\documentclass{article}

     \usepackage[final]{nips_2018}

\usepackage[utf8]{inputenc} 
\usepackage[T1]{fontenc}    
\usepackage{hyperref}       
\usepackage{url}            
\usepackage{booktabs}       
\usepackage{amsfonts}       
\usepackage{nicefrac}       
\usepackage{microtype}      
\usepackage{mathtools}
\usepackage{graphicx}
\usepackage{mwe}
\usepackage{subfig}
\usepackage{caption}

\title{Prediction of Small Molecule Kinase Inhibitors for Chemotherapy Using Deep Learning}

\author{%
  Niranjan Balachandar \\
  \texttt{niranja9@stanford.edu} \\
  \And
  Christine Liu \\
  \texttt{cliu99@stanford.edu} \\
  \AND
  Winston Wang \\
  \texttt{wwang13@stanford.edu} \\
}

\begin{document}

\maketitle

\begin{abstract}
The current state of cancer therapeutics has been moving away from one-size-fits-all cytotoxic chemotherapy, and towards a more individualized and specific approach involving the targeting of each tumor's genetic vulnerabilities.  Different tumors, even of the same type, may be more reliant on certain cellular pathways more than others. With modern advancements in our understanding of cancer genome sequencing, these pathways can be discovered. Investigating each of the millions of possible small molecule inhibitors for each kinase \textit{in vitro}, however, would be extremely expensive and time consuming. This project focuses on predicting the inhibition activity of small molecules targeting 8 different kinases using multiple deep learning models. We trained fingerprint-based MLPs and simplified molecular-input line-entry specification (SMILES)-based recurrent neural networks (RNNs) and molecular graph convolutional networks (GCNs) to accurately predict inhibitory activity targeting these 8 kinases.
\end{abstract}

\section{Introduction and Background}
Cancer continues to be one of the world's most common and deadly health problems. The American Cancer Society reports that approximately 1.8 million people will be diagnosed with cancer each year in the United States, and another 600,000 will succumb to the disease \cite{siegal2018statistics}. For patients with cancer, traditional treatment options including surgery, chemotherapy, and radiation therapy. Each of these treatments has their own set of side effects, with chemotherapy in particular having poor efficacy and nontrivial toxicity levels \cite{huang2017effects}.

Modern cancer therapies have started to shift away from this “one-size-fits-all” cytotoxic chemotherapy approach, and towards a more individualized targeting of specific tumors’ genetic vulnerabilities. As researchers continue to gain a better understanding of cancer genome sequencing, our understanding of how particular cellular pathways are related to tumorigenics of different types of cancer. In this paper, we focus on the inhibition of kinases, which are one of the most intensively pursued classes of proteins for cancer drug discovery. There are approximately 518 kinases encoded within the human genome, with approximately 30 distinct kinase targets currently being developed for Phase I clinical trial. Searching for small molecule inhibitors for each kinase would be extremely expensive and time consuming, but the utilization of machine learning techniques can expedite the process, therefore allowing for a faster and more accessible approach. 

\subsection{Types of Kinases}
In this paper, we focus on inhibitors of 8 particular protein kinases. They are cyclin-dependent kinase, epidermal growth factor receptor, glycogen synthase kinase-3 beta, hepatocyte growth factor receptor, MAP kinase p38 alpha, tyrosine-protein kinase LCK, tyrosine-protein kinase SRC, and vascular endothelial growth factor receptor 2. Cyclin-dependent kinases (CDK) are important for the regulation of various phases and transitions of the cell cycle. CDKs are often the target of tumorigenic signals that can lead cyclin D to form complexes in the G1/S transition phase of the eukaryotic cell cycle. Epidermal growth factor receptors (EGFR) are activated by the binding of ligands to the extra-cellular domain, which leads to signal transduction and activation of various pathways involved in cell proliferation and survival. EGFR pathway dysregulation is found in multiple human cancers. For example, overexpression of EGFR is linked to ~80 percent of non-small cell lung cancer, and mutation of EGFR is linked to ~20 percent of non-small cell lung cancer \cite{kannaiyan}. Glycogen synthase kinase-3 beta (GSK3$\beta$) is a protein kinase that has established as a tumor suppressor or promoter in various cancers such as skin, oral, larynx, breast and other types of cancer. GSK3$\beta$ is active in resting cells, and is regulated by various upstream kinases which can cause rapid inhibition or promotion \cite{Mishra}.  Hepatocyte growth factor receptors (HGF) are often linked to tumor progression in several cancers as they have a major role in embryonic organ development and adult organ regeneration. \cite{cecchi} MAP kinase p38 alpha (MAPK) are stress-activated serine/threonine-specific kinases. MAPK is linked to diverse cellular functions such as gene expression and cell death. \cite{olson2004} Tyrosine-protein kinase (LCK) regulates cell growth, migration, and proliferation. Upregulation of LCK is often correlated with breast cancer \cite{kannaiyan}. Tyrosine-protein kinase (SRC) is also involved in growth, migration, and proliferation, but overexpression is linked to colon cancers \cite{kannaiyan}. Vascular endothelial growth factor receptor 2 (VEGFR-2) is involved in mediating pro-angiogenic signals. The greater signal intensity for VEGFR-2 is correlated to human cancer tumors \cite{smith2010}.

\section{Related Work}
Novel approaches to oncology drug discovery has been largely spurred by the new understanding of how abnormal protein kinase activity has been linked to the development and onset of a variety of diseases. Different tumors, even of the same type, may be more reliant on certain cellular pathways more than others, which can be discovered through cancer genome sequencing. Previous research has been done in order to characterize the structural basis of kinase inhibitor selectivity, and identify potential kinases that are involved in cancer mechanisms, but there is a lacking of research done to predict potential inhibitors. Once kinases have been identified, the IC-50 values, SMILES codes, connectivity, and other features can be utilized in conjunction with deep learning models to predict the activity of small molecule inhibitors \cite{zhang2009inhibitors}.

Briem et al. developed a model using artificial neural networks (ANNs) in order to distinguish between kinase inhibitors and non-inhibitors for 8 kinases \cite{briem}. Their data comprised of 7759 training kinase inhibitor compounds, with 565 of the molecules classified as "actives" and 7194 classified as "inactives". A molecule was defined as "active" is it had an $IC_{50}\leq=10\mu M$ in at least 1 out of 8 kinase assays performed in-house. There was also a separate validation test set of 504 molecules, with 204 "actives" and 300 "inactives". Briem et al. applied a standard feed-forward neural network through the implementation of the TSAR software package \cite{tsar}. They generated 13 models, and all layers were completely connected. By training on their binary dataset and using the majority votes from the 13 models, they were able to achieve an accuracy score of 0.88, a precision score of 0.84, a recall score of 0.88, and a F1 score of 0.86. Briem et al. also developed a model based on support vector machines (SVM), with an accuracy score of 0.88, a precision score of 0.86, a recall score of 0.85, and a F1 score of 0.86. The work done by Briem et al. was the first time that prediction tools were used to try and capture essential features of kinase inhibitor molecules, as opposed to druglike molecules that only target particular families. The idea of personalized kinase inhibition is novel, and Briem et al. were one of the first groups to propose comparing and identifying features of the kinase inhibitors, rather than predict successful drugs. 

More recently, research groups have been interested in learning how to represent the molecular structure information in a more robust way. While common fingerprinting methods are generally successful at representing small molecules well, they are non-differentiable and cannot adapt to emphasize molecular structure aspects differently based on data given. The Pande Group at Stanford University has been focusing on a new representation of small molecules as undirected graphs of atoms.  Graph convolutional neural networks aim to featurize molecules in a differentiable way, so that the way the molecule is represented can change based on the data and task given. Riley et al. highlights the flexibility of the graph convolution architecture, despite the model performing similar to fingerprint-based approaches \cite{graph_conv_intro}. 

In our study, we expanded on the work done by Briem et al. by using specific structural features of the inhibitors, such as SMILES codes and molecular fingerprints, in order to predict which molecules could serve as inhibitors for various kinases. In addition, we also wanted to explore the GCN architecture proposed by Pande and Riley et al. Using the various types of data available, we were able to try a variety of models and approaches and achieve consistent results.

\subsection{Dataset}
For this project, we utilize the Kaggle Cancer Inhibitors Dataset \cite{kaggle}.  It consists of labeled inhibitors (positive class) and non-inhibitors (negative class) for each of eight kinases. Statistics for the Kaggle data for each kinase are summarized in Table \ref{table:kaggle_stats}. For each inhibitor, the Kaggle dataset includes a ChemBL accession ID and binary inhibition label. The vast majority of molecules have labels for only one of the eight kinases, so it is impractical to perform multitask learning - we train separate models for each of the eight kinases. We get ChemBL IDs and labels from Kaggle, but we need feature representations of the molecules to feed into deep learning models. Thus we use the ChemBL database to gather SMILES strings and the Python RDKit package to generate fingerprints for each molecules - the SMILES strings and fingerprints will serve as feature representations for the small molecules 
\begin{table}[htbp!] 
\begin{center}
\begin{tabular}{l |r |r |r}
Kinase & Total & Inhibitors & Non-Inhibitors \\
\hline
CDK2 & 1635 & 1039 & 596  \\
EGFR\textunderscore ERB1 & 5176 & 3554 & 1622 \\
GSK3B & 1950 & 1434 & 516 \\
HGFR & 2145 & 1830 & 315 \\
MAP\textunderscore K\textunderscore P38A & 3597 & 3080 & 517 \\
TPK\textunderscore LCK & 1809 & 1302 & 507 \\
TPK\textunderscore  SRC & 2972 & 1891 & 1081 \\
VEGFR2 & 5123 & 4089 & 1034  
\end{tabular}
\caption{Total, positive (inhibitors), and negative (non-inhibitor) samples for each kinase in Kaggle dataset}
\label{table:kaggle_stats}
\end{center}
\end{table}
\subsection{SMILES}
SMILES stands for "simplified molecular-input line-entry system" and is a specification in form of a line notation for describing the structure of chemical species using short strings that can be understood for data processing \cite{smiles}. SMILES strings are unique for each structure. An overview of the SMILES conversion algorithm starts with the conversion of the SMILES to an internal representation of the molecular structure and then examines structure and produces a unique SMILES string. SMILES codes are generated by first breaking the cycles of the molecules, and then writing the branches off a main backbone. The unique codes are dependent on the bonds chosen to break cycles, on starting atom used for depth-first traversal, charge, and order in which branches are listed when encountered. We obtain a unique SMILES string for each small molecule by querying the ChemBL database using the corresponding ChemBL accession ID provided in the Kaggle dataset.

\subsection{Fingerprints}
We also choose to represent each small molecule as fingerprints - a unique bit vector corresponding to the molecule that encode 2D, 3D, and chemical properties. In particular, we use three common fingerprints - Atom-Pair, Morgan, and Topological Torsion. Atom pairs are a substructure within each molecule that is defined by the atomic environment and shortest subpath between all pairs of atoms in the topological representation of a molecule\cite{atom_pair}.  Atom-pair fingerprints are generated based on correlations between different atom-pair structures and biological activity.  Morgan fingerprints, or Extended Connectivity fingerprints, are circular fingerprints generated from the topological representation of the molecule\cite{ecfp}.  These fingerprints represent the molecule by the circular neighborhoods around each atom.  Finally, the Topological Torsion fingerprint is based on sets of four consecutively connected non-hydrogen atoms, which represent a torsion angle\cite{topological_torsion}. We generate each of these three fingerprints using the Python RDKit package by feeding in SMILES string. Each fingerprint is a bit vector of length 2048, and for each molecule we concatenate the three fingerprints into a bit vector of length 6144.

\subsection{Training, Validation, and Testing Splits}
Initially, we randomly split the small molecules for each kinase into training, validation, and testing with a 0.7:0.15:0.15 ratio. However, in order to make sure that the models do indeed learn novel chemical features rather than just learn to classify related molecules from the training set, we create harsher splits using the following approach. For each kinase we cluster the small molecules based on their concatenated fingerprint vectors into 7 clusters using K-means, randomly select 5 of these clusters for training, and divide the remaining molecules into validation and testing with a 0.5:0.5 ratio. We include results from both splits. 

\subsection{Baseline MLP}\label{section:mlp}
We will first directly feed the concatenated fingerprint feature representation into a multilayer perceptron (MLP) with a two hidden dense layers of size 128, each with ReLU activation. Before the output layer, we include a dropout layer for regularization. The output layer is a dense layer of size 2 - corresponding to the two possible labels - with softmax activation for a probability prediction for each label. The MLP architecture is summarized in Figure \ref{fig:model_mlp}. We use binary cross entropy as the loss function, as given in the following equation:
$$ J(\theta) = -\sum_{i=1}^N (y_i\log(\hat{y_i})+(1-y_i)\log(1-\hat{y_i}))$$
where $J$ is the loss, $\theta$ is the current set of model parameters, $N$ is the total number of data samples, $y_i$ is true label (0 or 1) for data sample $i$, and $y_i$ is predicted probability output by the model that the label for data sample $i$ is 1. Adam is a gradient descent algorithm with an adaptive learning rate that in practice yields quicker model convergence than vanilla stochastic gradient descent, so we use Adam to optimize our model \cite{kingma2014adam}. We set the initial learning rate to $0.001$, minibatch size to 32, and dropout probability to $0.5$.

\subsection{Recurrent Neural Network}\label{section:rnn}
\begin{figure}[htbp!]   
\centering
\subfloat[]{\includegraphics[width=0.25\textwidth]{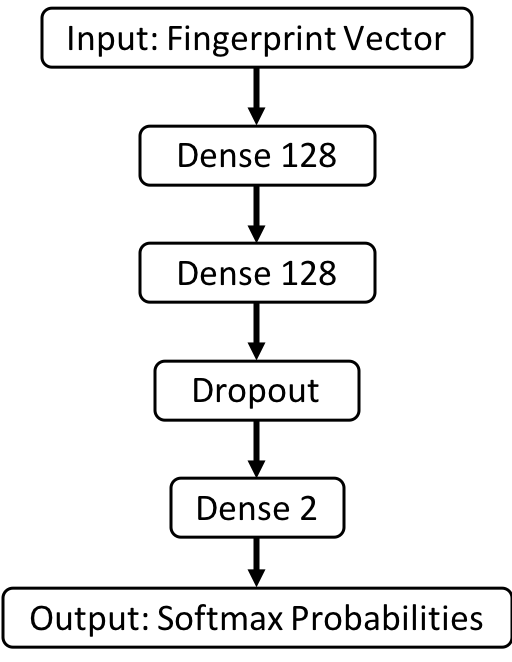}\label{fig:model_mlp}}
\qquad
\subfloat[]{\includegraphics[width=0.25\textwidth]{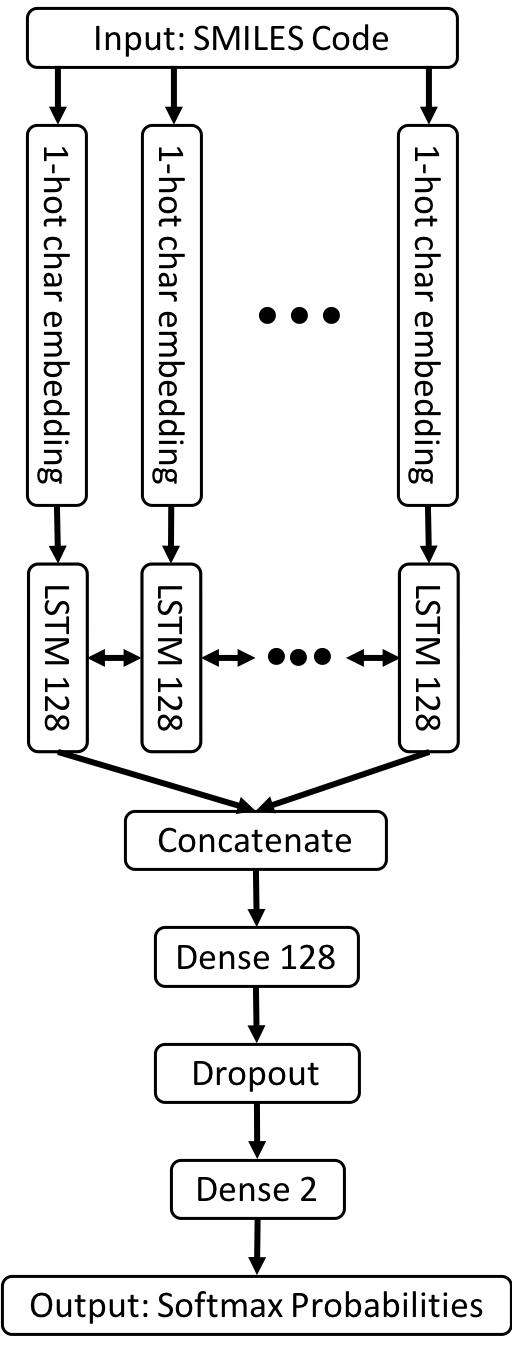}\label{fig:model_rnn}}
\caption{(a) MLP architecture and (b) RNN architecture}
\label{fig:models}
\end{figure}

To take advantage of the molecular structure and atomic connectivity of the small molecules, we apply a Recurrent Neural Network (RNN) on the SMILES code of the small molecules. We pad the SMILES codes of each molecule within a certain kinase dataset with the null character `!' so that each SMILES code for a particular kinase has equal length for batch train (the codes for different kinases may have different length). We then convert the SMILES codes to 1-hot vector embeddings. This is done by first enumerating all the characters that appear in all the SMILES codes for a particular kinase in an arbitrary order. Then, each character of the SMILES code is converted to a 1-hot vector that is 1 at the vector index corresponding to the enumeration ID of that character, and 0 at all other indices. These 1-hot vectors are then fed into a bidirectional Long-term Short-term Memory Network (LSTM) with 128 hidden units. We use a LSTM because it mitigates the vanishing gradient problem of standard RNNs \cite{hochreiter1997long}. We use a bidirectional LSTM because the atomic connectivity of a molecule is unordered. The output of the first and last LSTM layers are concatenated into a vector of size 256, and this vector is fed a dense layer of size 128 with ReLU activation. Before the output layer, we include a dropout layer for regularization. The output layer is a dense layer of size 2 with softmax activation for a probability prediction for each label. The RNN architecture is summarized in Figure \ref{fig:model_rnn}. Like for the MLP, we use binary cross entropy as the loss function and Adam for gradient descent. We set the initial learning rate to $0.001$, minibatch size to 32, and dropout probability to $0.5$. Because of the exploding gradient problem, we clip the gradient norm to 5. 

\subsection{Fused Models}
\begin{figure}[htbp!]   
\centering
\subfloat[]{\includegraphics[width=0.42\textwidth]{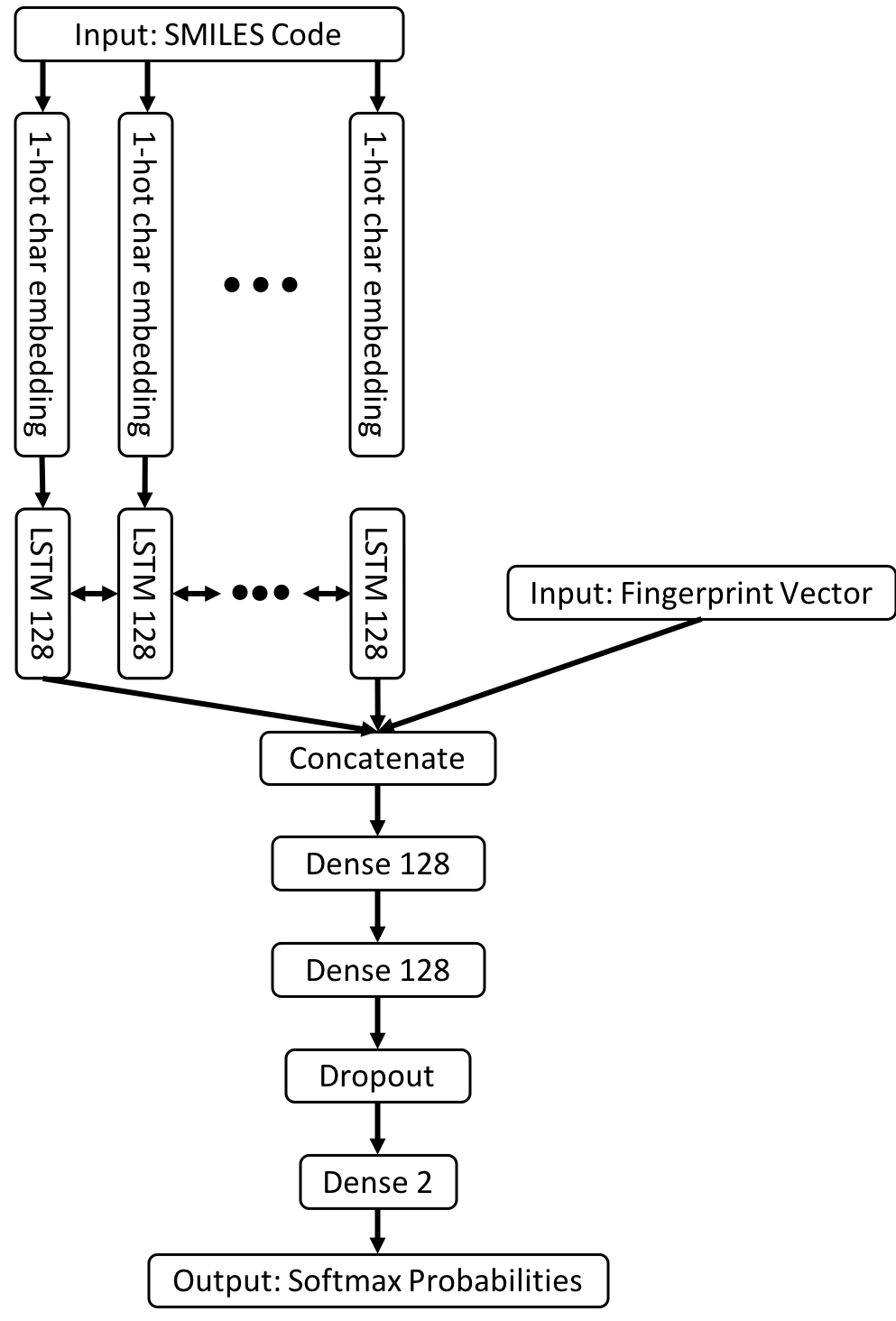}\label{fig:fused_model_early}}
\qquad
\subfloat[]{\includegraphics[width=0.42\textwidth]{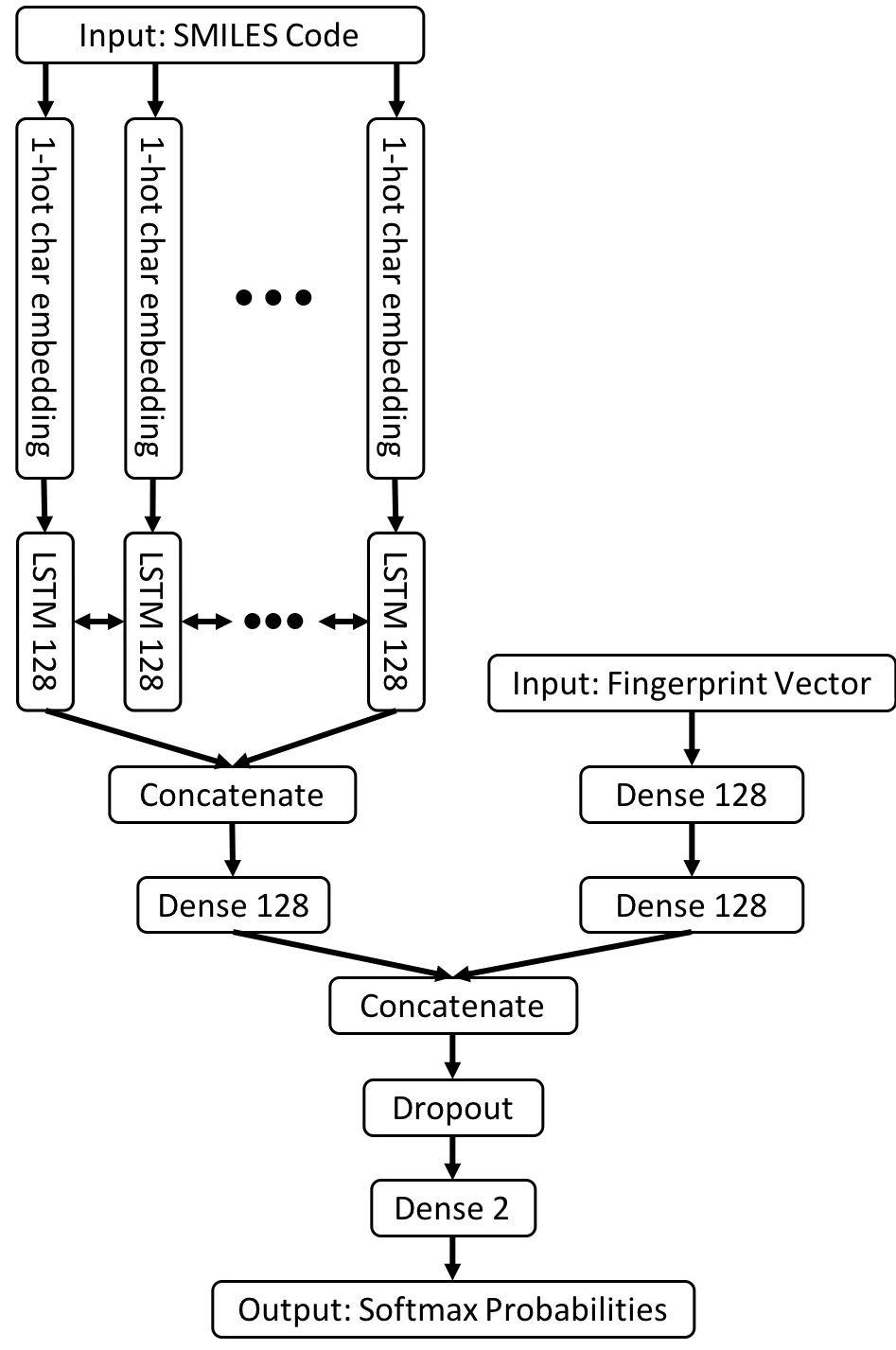}\label{fig:fused_model_late}}
\caption{(a) RNN+MLP early fusion architecture and (b) RNN+MLP late fusion architecture}
\label{fig:fused_models}
\end{figure}

To leverage both the chemical features obtained from molecular fingerprints and the structural features of the SMILES codes, we implement a number of approaches to fuse the fingerprint-based MLP with the SMILES-based RNN. The first approach is early fusion, where the outputs of the first and last pretrained LSTM outputs are concatenated with the fingerprint feature vector to construct a vector of size 6400. This concatenated vector is then fed as input into the MLP as described in Section \ref{section:mlp}. The early fusion model is illustrated in Figure \ref{fig:fused_model_early}, and all layers are fine-tuned during training. The second approach is late fusion, where the output of the final dense layer of size 128 from the pretrained RNN as described in section \ref{section:rnn} is concatenated with the output of the final dense layer of size 128 from the pretrained MLP as described in section \ref{section:mlp} to produce a vector of size 256. This vector is then fed through a dropout layer for regularization, and a final output layer of size 2 with softmax activation to generate probability predictions. The late fusion model is illustrated in Figure \ref{fig:fused_model_late}, and all layers are fine-tuned during training. Like for the RNN, we use binary cross entropy as the loss function and Adam for gradient descent. We set the initial learning rate to $0.001$, minibatch size to 32, dropout probability to $0.5$, and gradient clipping norm to 5. The final fused model is a simple ensemble between the MLP and RNN where the predicted probabilities from the two models are averaged to produce the prediction for the ensemble model.

\subsection{Graph Convolutional Neural Networks}
In order to perform deep learning on molecules, we must first find a way to featurize the molecule into a vector, which can then be the input into any neural network.  Common methods to featurized molecules, such as Extended Connectivity Fingerprints (ECFP) and other fingerprinting methods, are able to represent small molecules well, however, they cannot adapt to the given task, as they are non-differentiable.  Graph convolutional neural networks aim to featurize molecules in a differentiable way, so that the way the molecule is represented can change based on the task and dataset at hand.  

\begin{figure}
    \centering
    \includegraphics[width=0.4\textwidth]{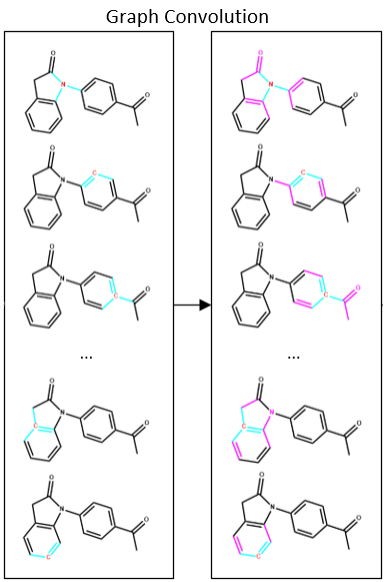}
    \caption{After each convolution, the feature vectors for each atom represent a weighted average of its neighbors \cite{chemi-net}}
    \label{fig:neighbor_flow}
\end{figure}

Essentially, a graph convolution layer, similar to a convolutional layer, represents each node as a combination of its neighbors, as shown in Figure \ref{fig:neighbor_flow}.  This is accomplished by feeding both atom features (\textit{n}-dimensional vectors for each atom) and pair features (\textit{n}-dimensional vectors for each pair of atoms) through Weave modules in series\cite{graph_conv_intro}.  Weave modules combine these atom and pair features (denoted $A$ and $P$) together to generate another set of atom and pair features, which can then be fed into another Weave module.  The architecture of a Weave module is shown in Figure \ref{fig:weave}, and the definitions of each of the operations is shown below:
\begin{align*}
    A\rightarrow A: A^{y} &= f(A^x)\\
    P\rightarrow P: P^{y} &= f(x)\\
    A\rightarrow P: P^{y}_{a, b} &= g(v_1, v_2)\\
    v_1 &= f(A^x_a, A^x_b)\\
    v_2 &= f(A^x_b, A^x_a)\\
    P\rightarrow A: A^{y}_a &= g(v_1, v_2, v_3, \ldots)\\
    v_1 &= f(P^x_{a, b})\\
    v_2 &= f(P^x_{a, c})\\
    &\vdots
\end{align*}
where $f$ is an arbitrary, trainable function and $g$ is an arbitrary commutative function.

For this project, we represented each molecule as an atom feature vector and adjacency graph.  These two matrices were passed through four stacked Weave modules with a max atom size of 100, and fed the output through a single hidden layer of size 16.  We then utilized sigmoid cross-entropy loss to train.  As with the other models, we used the Adam optimizer. Finally, we also ensemble the GCN with the RNN and MLP by taking the average predicted probabilities across all three models.

\begin{figure}
    \centering
    \includegraphics[width=0.3\textwidth]{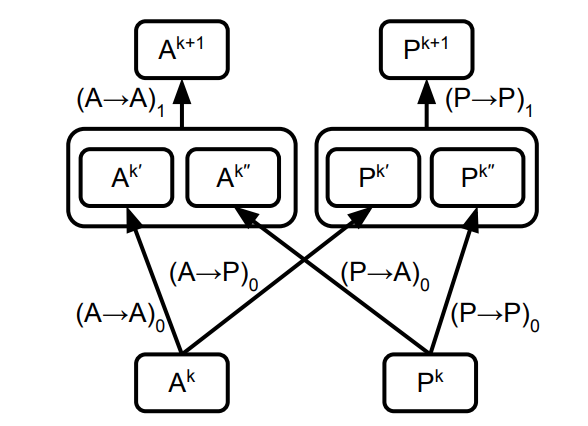}
    \caption{Graphic displaying the operations in a Weave module \cite{graph_conv_intro}}
    \label{fig:weave}
\end{figure}

\section{Results}
\begin{figure}[htbp!]
\begin{minipage}{.5\linewidth}
\centering
\subfloat[]{\label{fig:roc_a}\includegraphics[scale=.5]{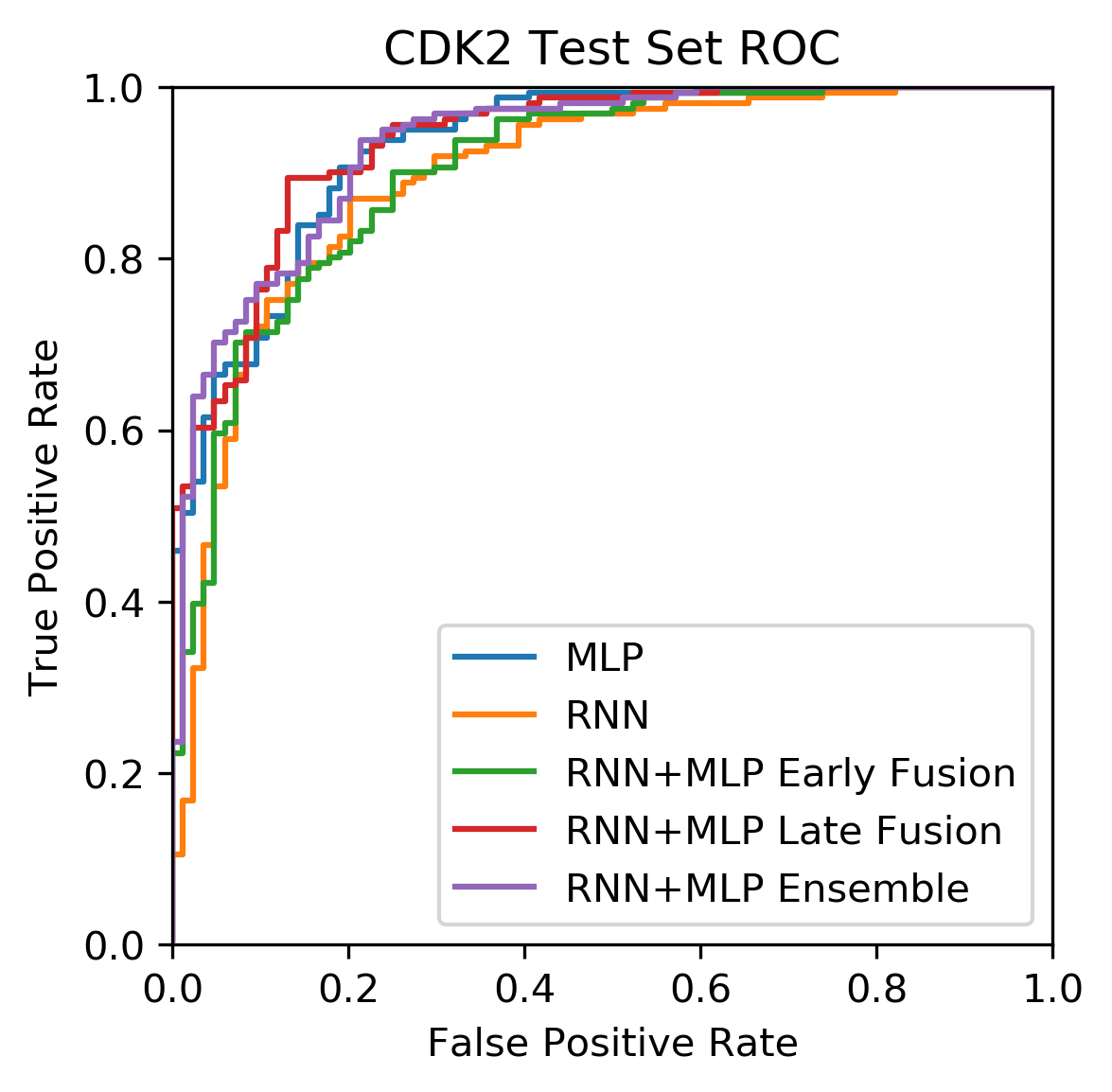}}
\end{minipage}%
\begin{minipage}{.5\linewidth}
\centering
\subfloat[]{\label{fig:roc_b}\includegraphics[scale=.5]{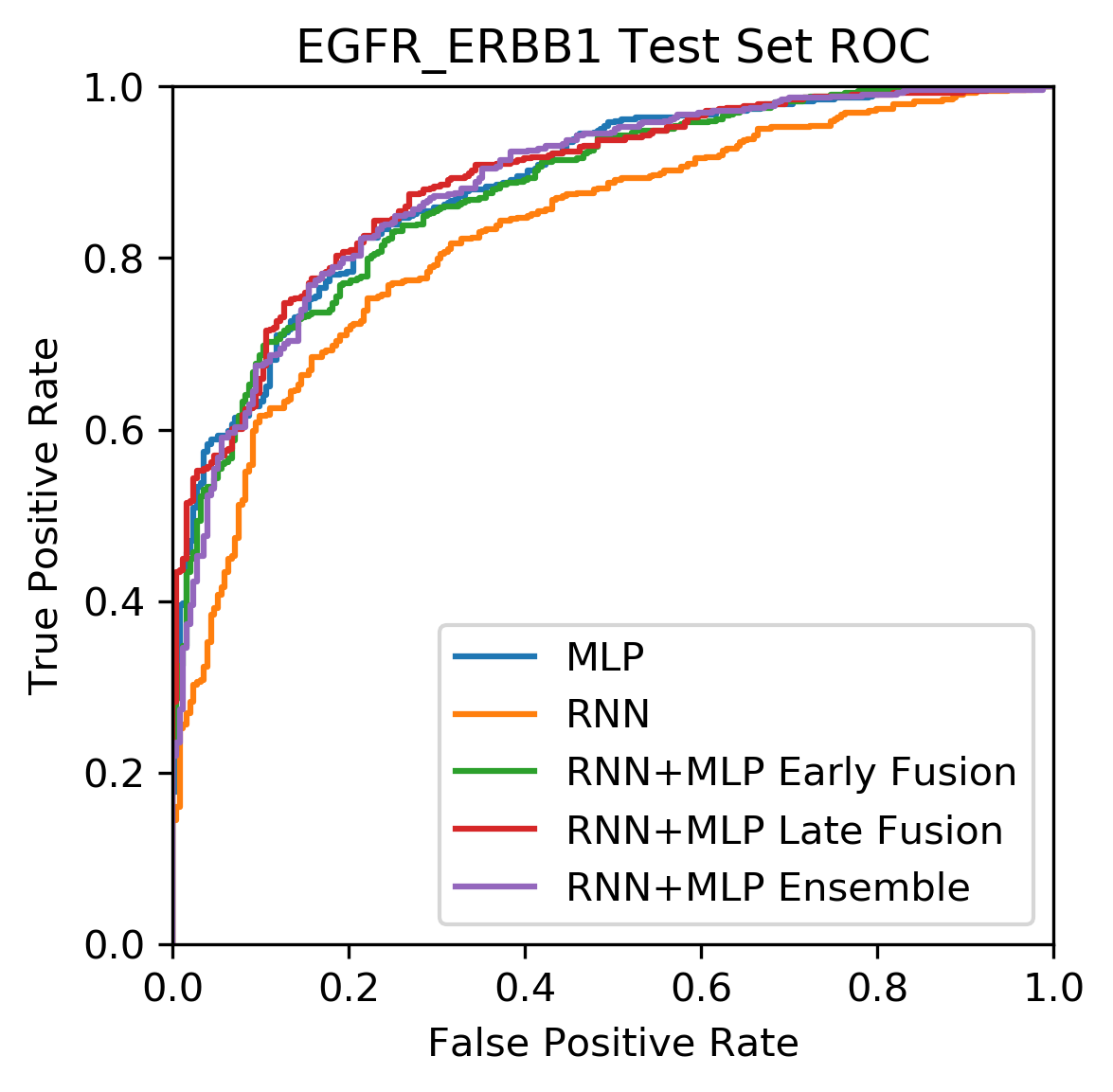}}
\end{minipage}\par\medskip
\centering
\begin{minipage}{.5\linewidth}
\centering
\subfloat[]{\label{fig:roc_c}\includegraphics[scale=.5]{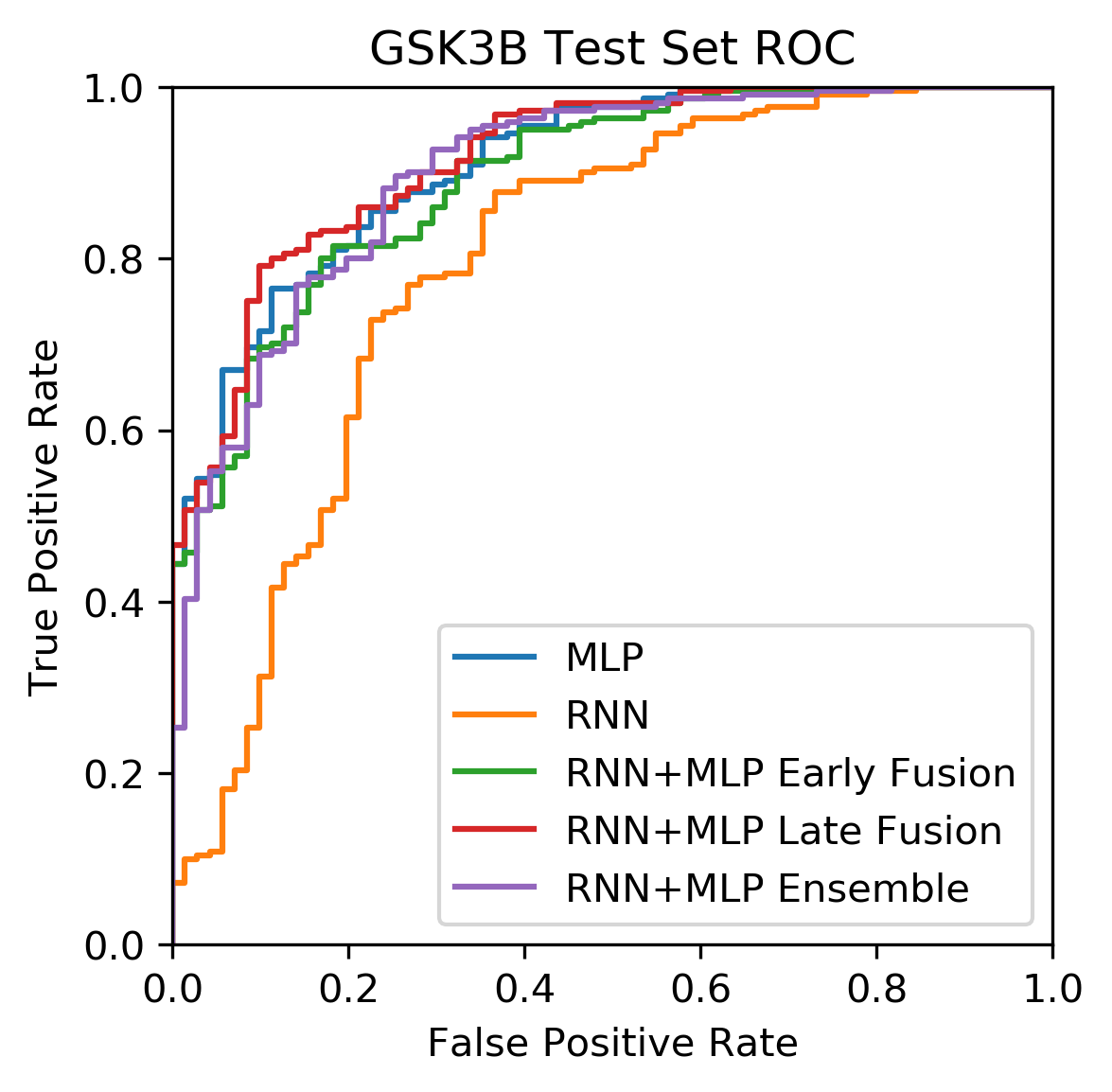}}
\end{minipage}%
\begin{minipage}{.5\linewidth}
\centering
\subfloat[]{\label{fig:roc_d}\includegraphics[scale=.5]{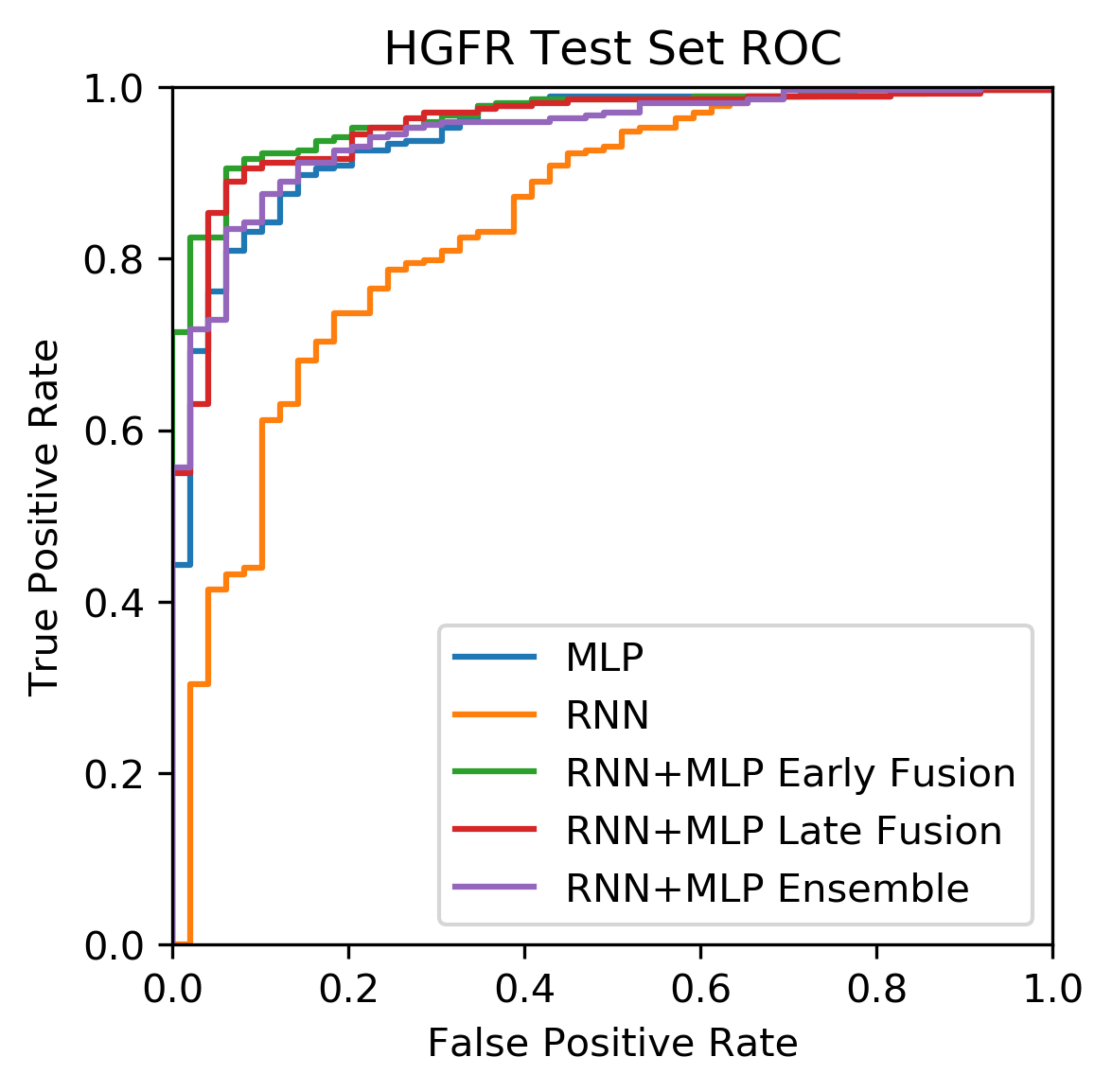}}
\end{minipage}\par\medskip
\begin{minipage}{.5\linewidth}
\centering
\subfloat[]{\label{fig:roc_e}\includegraphics[scale=.5]{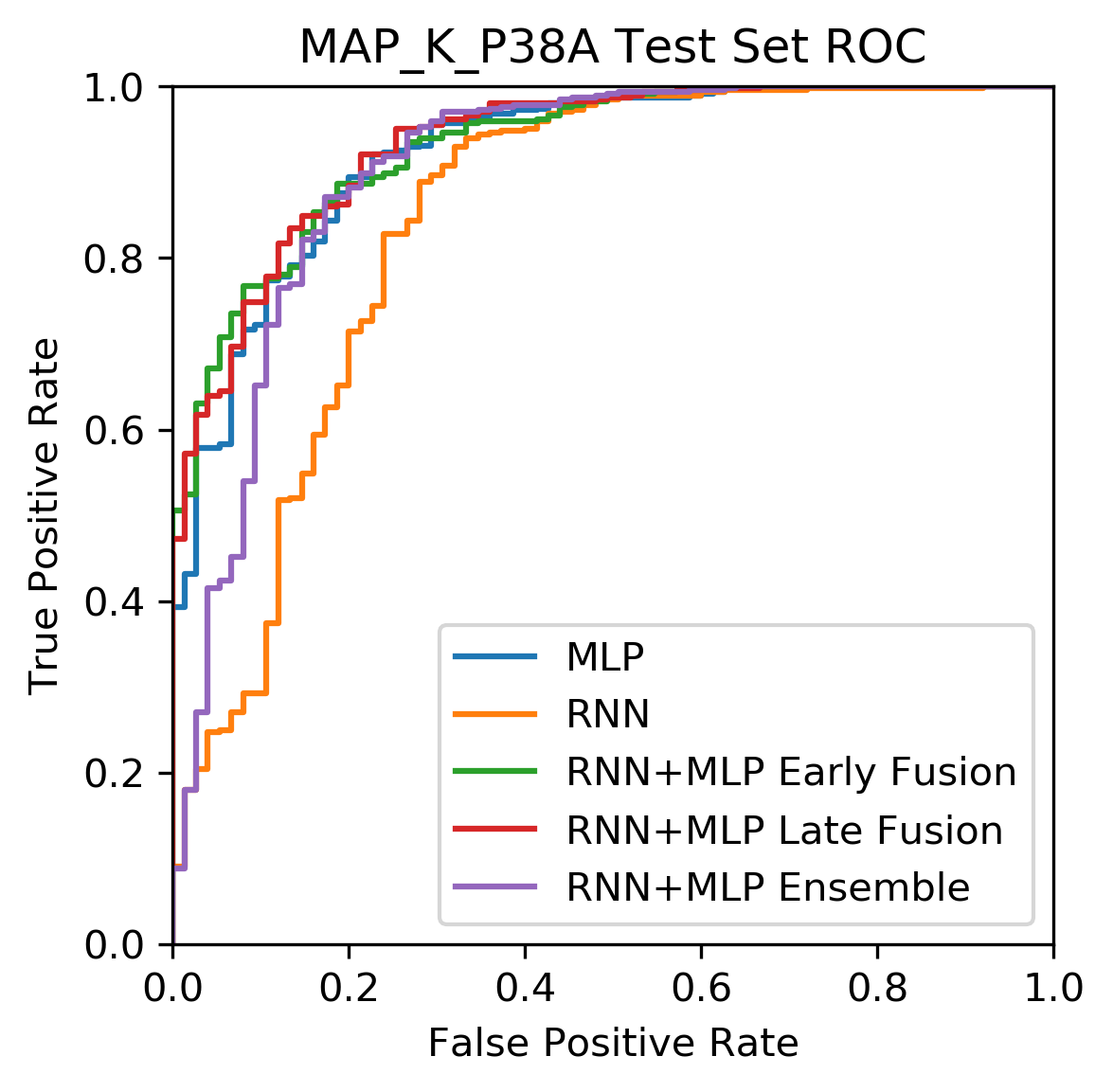}}
\end{minipage}%
\begin{minipage}{.5\linewidth}
\centering
\subfloat[]{\label{fig:roc_f}\includegraphics[scale=.5]{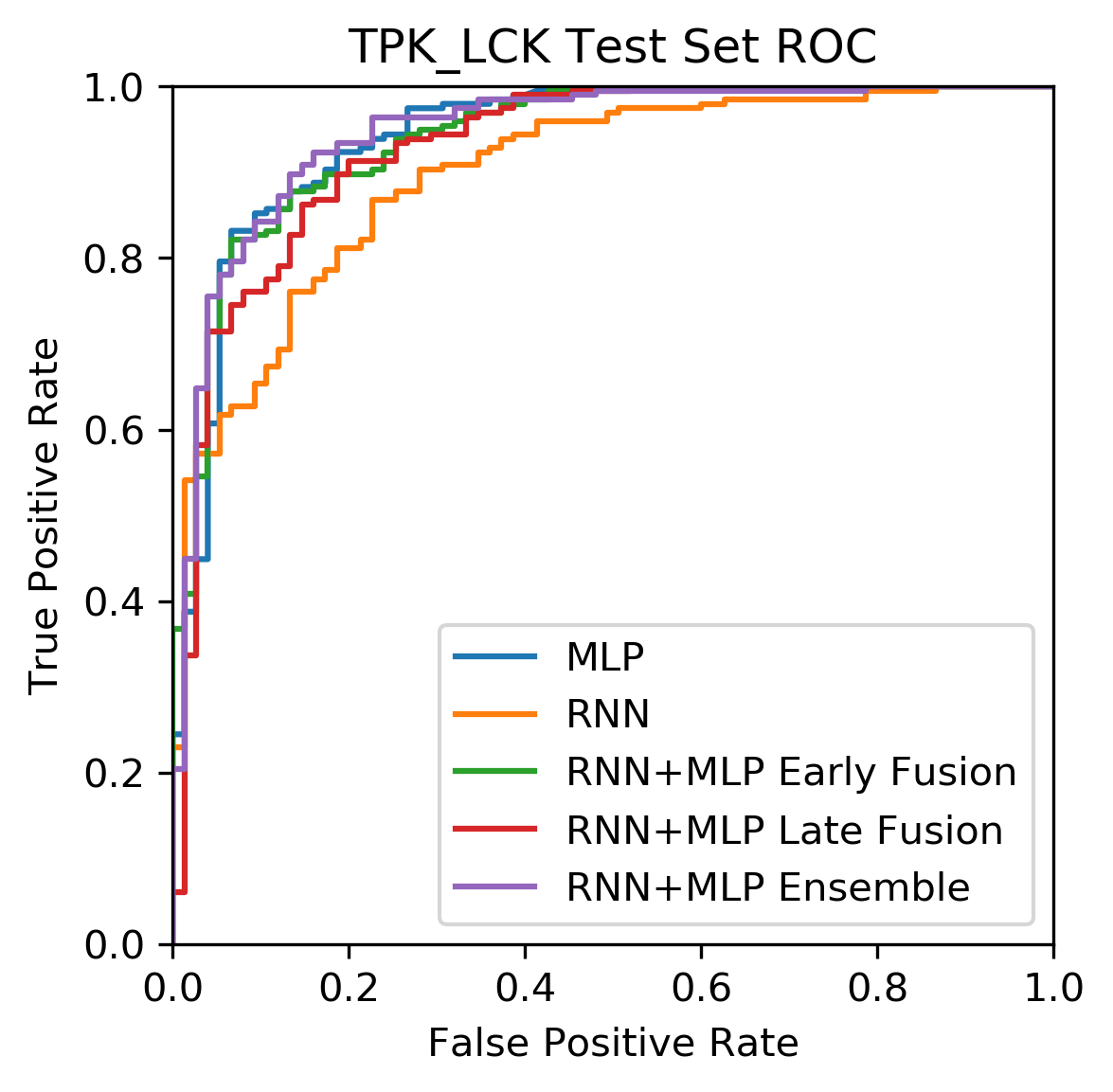}}
\end{minipage}\par\medskip
\begin{minipage}{.5\linewidth}
\centering
\subfloat[]{\label{fig:roc_g}\includegraphics[scale=.5]{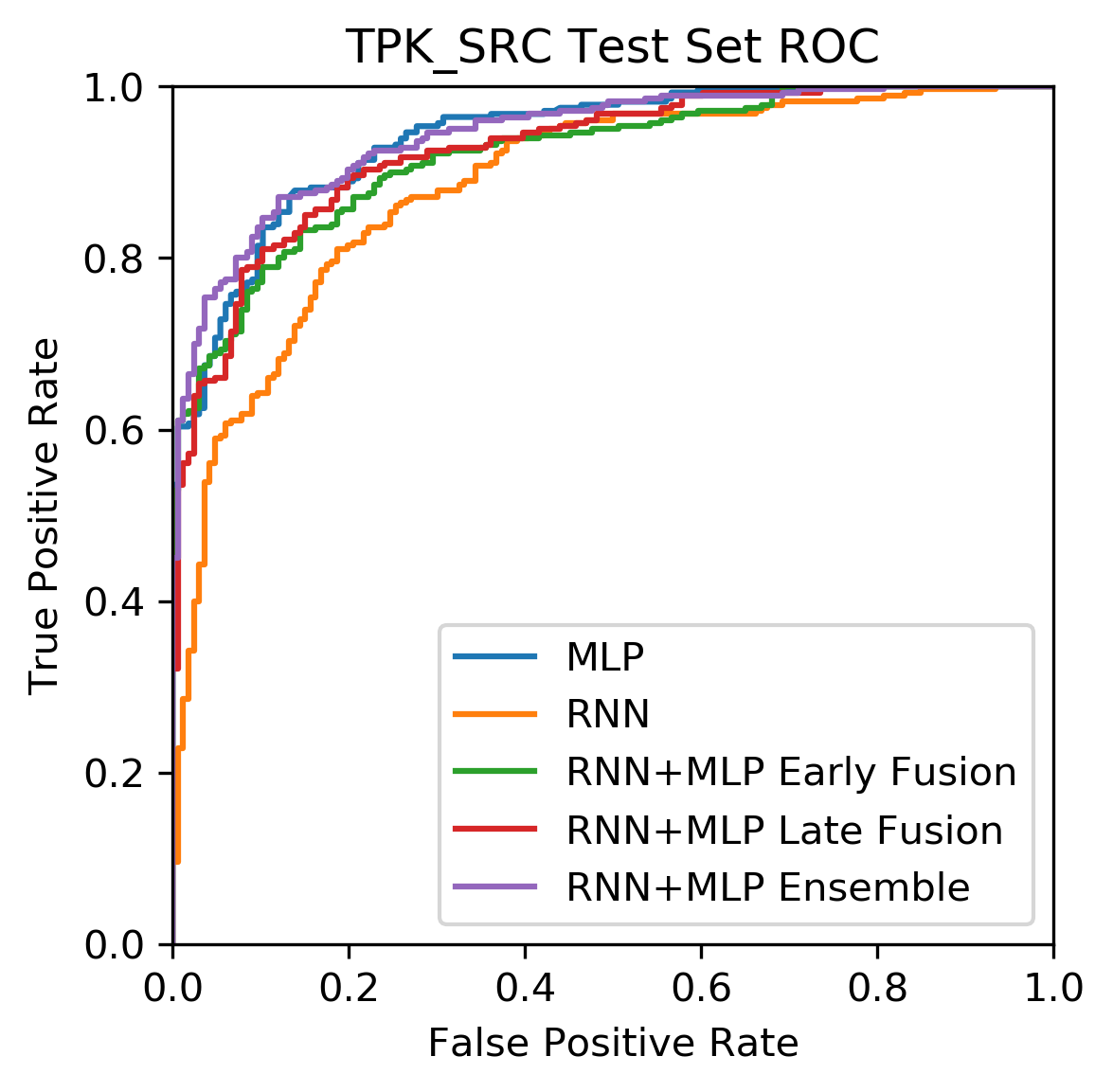}}
\end{minipage}%
\begin{minipage}{.5\linewidth}
\centering
\subfloat[]{\label{fig:roc_h}\includegraphics[scale=.5]{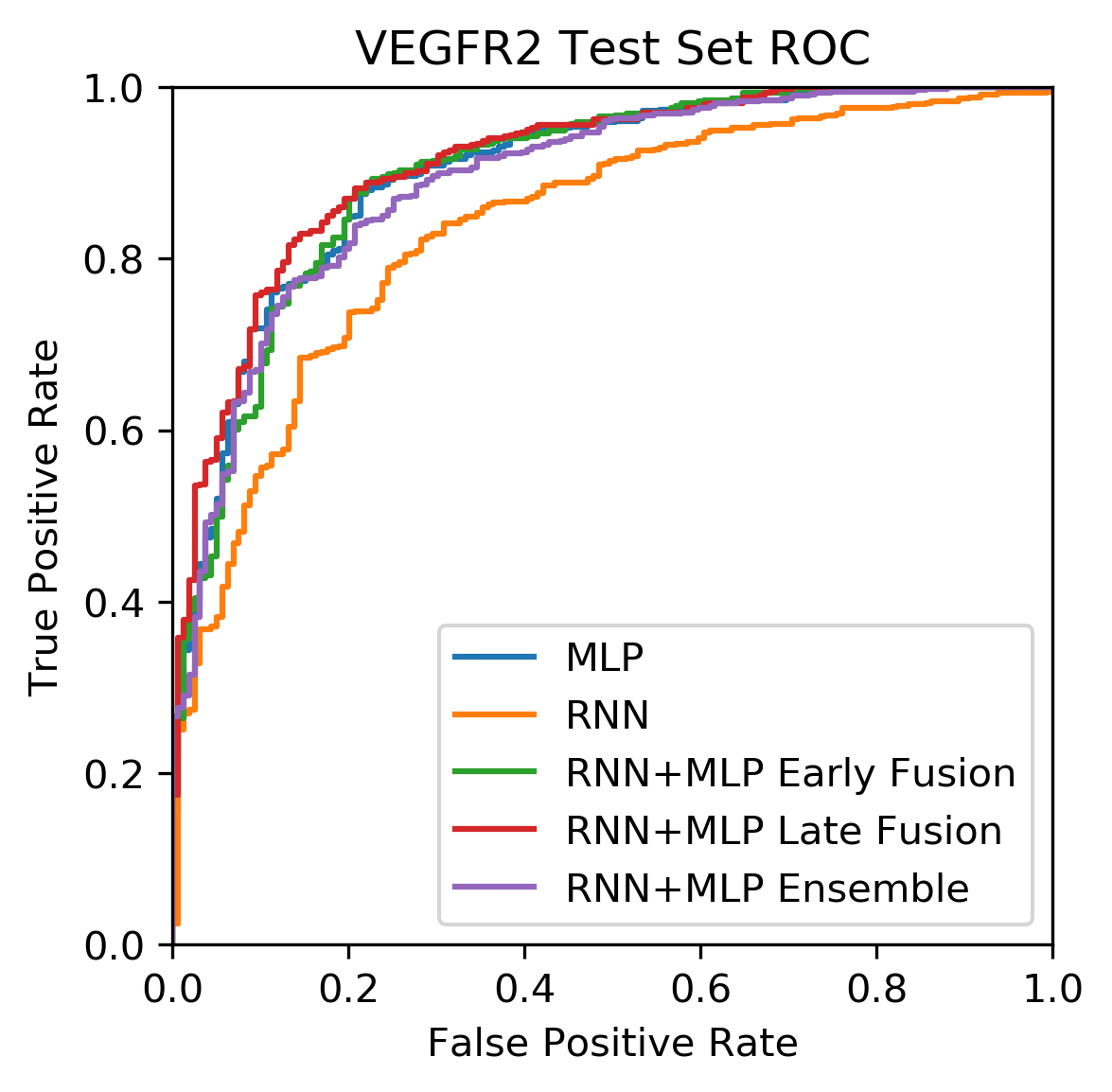}}
\end{minipage}
\caption{Testing set ROCs for each model for each of the 8 kinases using random splits.}
\label{fig:roc_random}
\end{figure}

\begin{figure}[htbp!]
\begin{minipage}{.5\linewidth}
\centering
\subfloat[]{\label{fig:prc_a}\includegraphics[scale=.5]{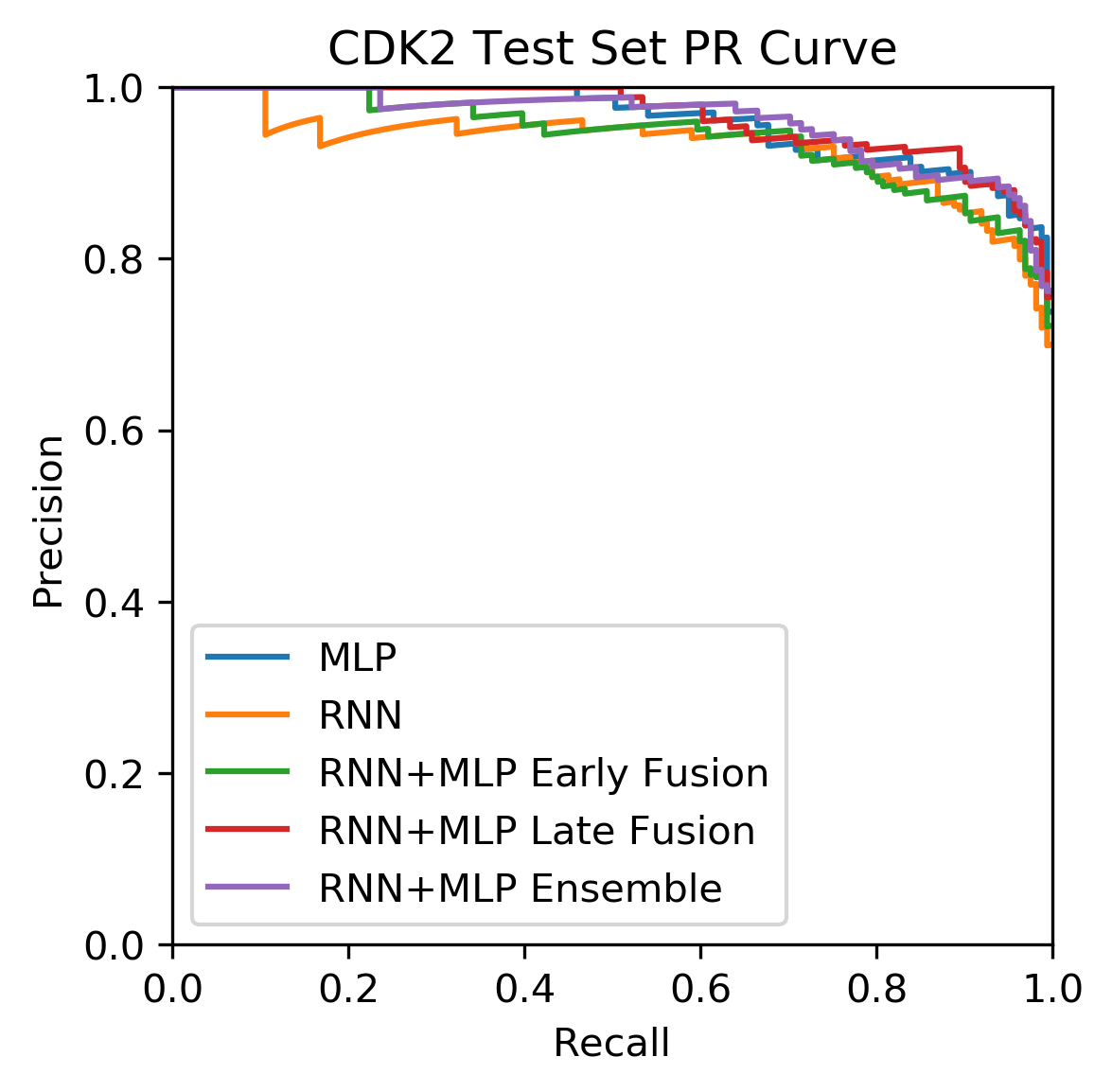}}
\end{minipage}%
\begin{minipage}{.5\linewidth}
\centering
\subfloat[]{\label{fig:prc_b}\includegraphics[scale=.5]{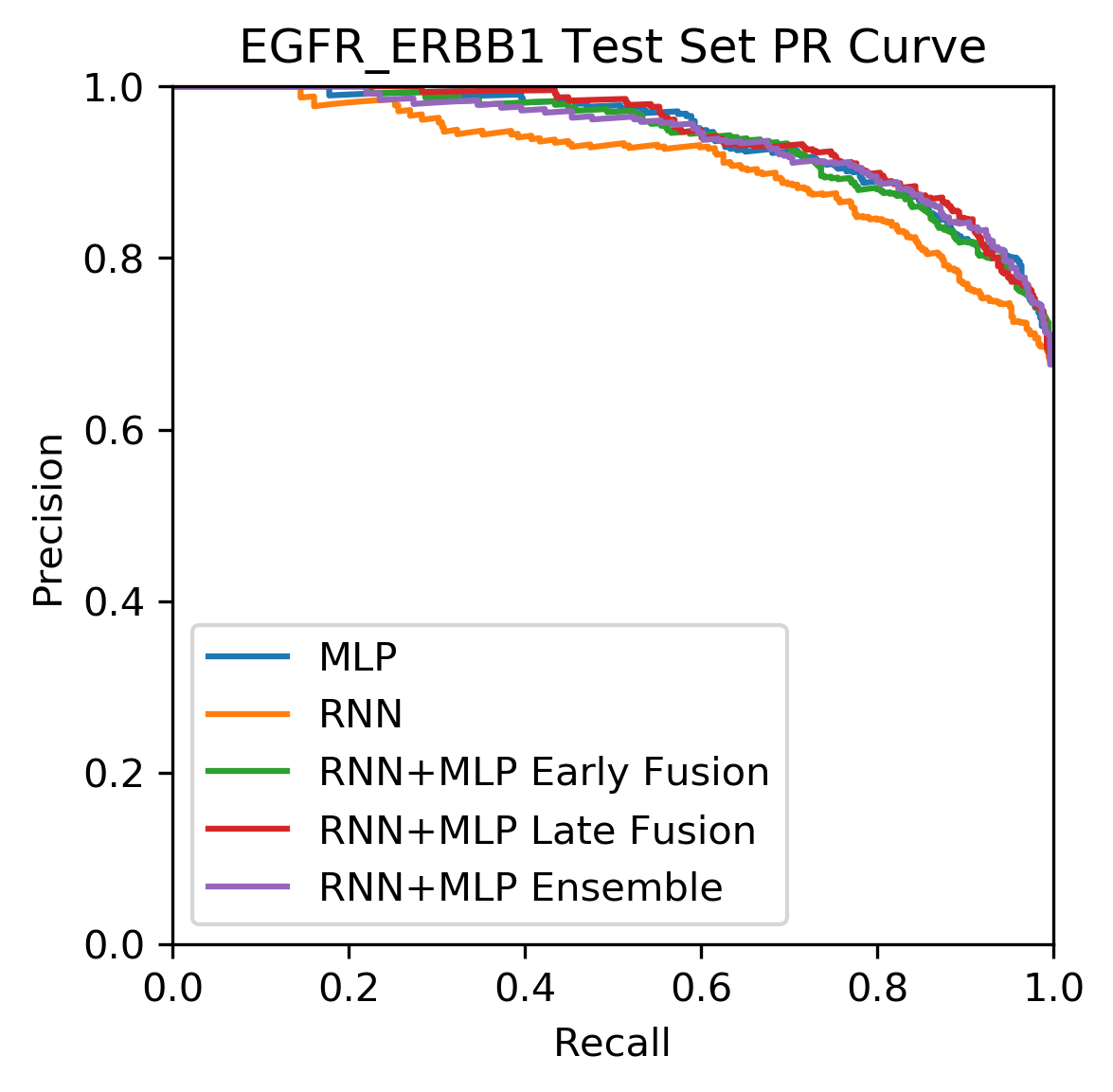}}
\end{minipage}\par\medskip
\centering
\begin{minipage}{.5\linewidth}
\centering
\subfloat[]{\label{fig:prc_c}\includegraphics[scale=.5]{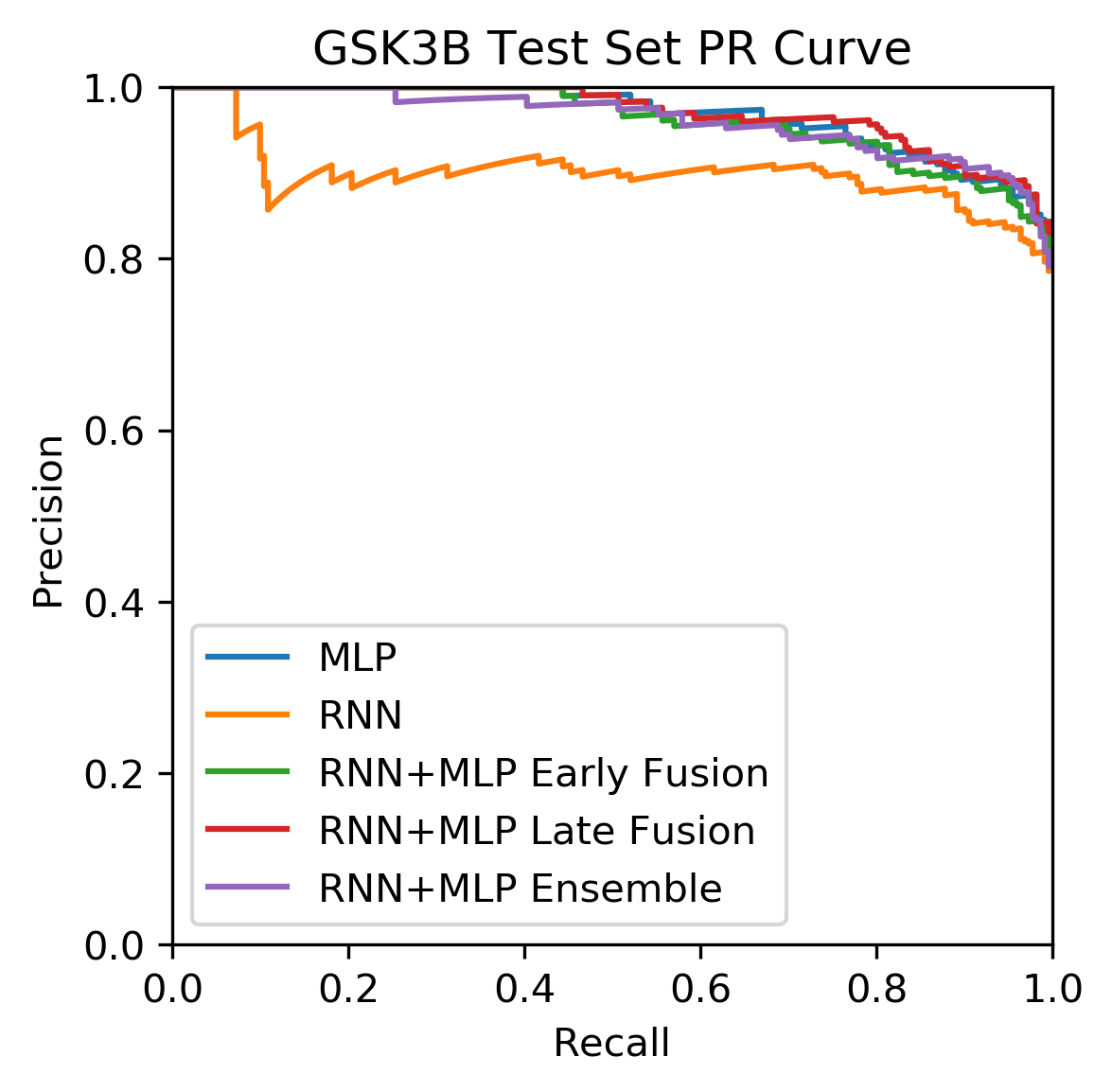}}
\end{minipage}%
\begin{minipage}{.5\linewidth}
\centering
\subfloat[]{\label{fig:prc_d}\includegraphics[scale=.5]{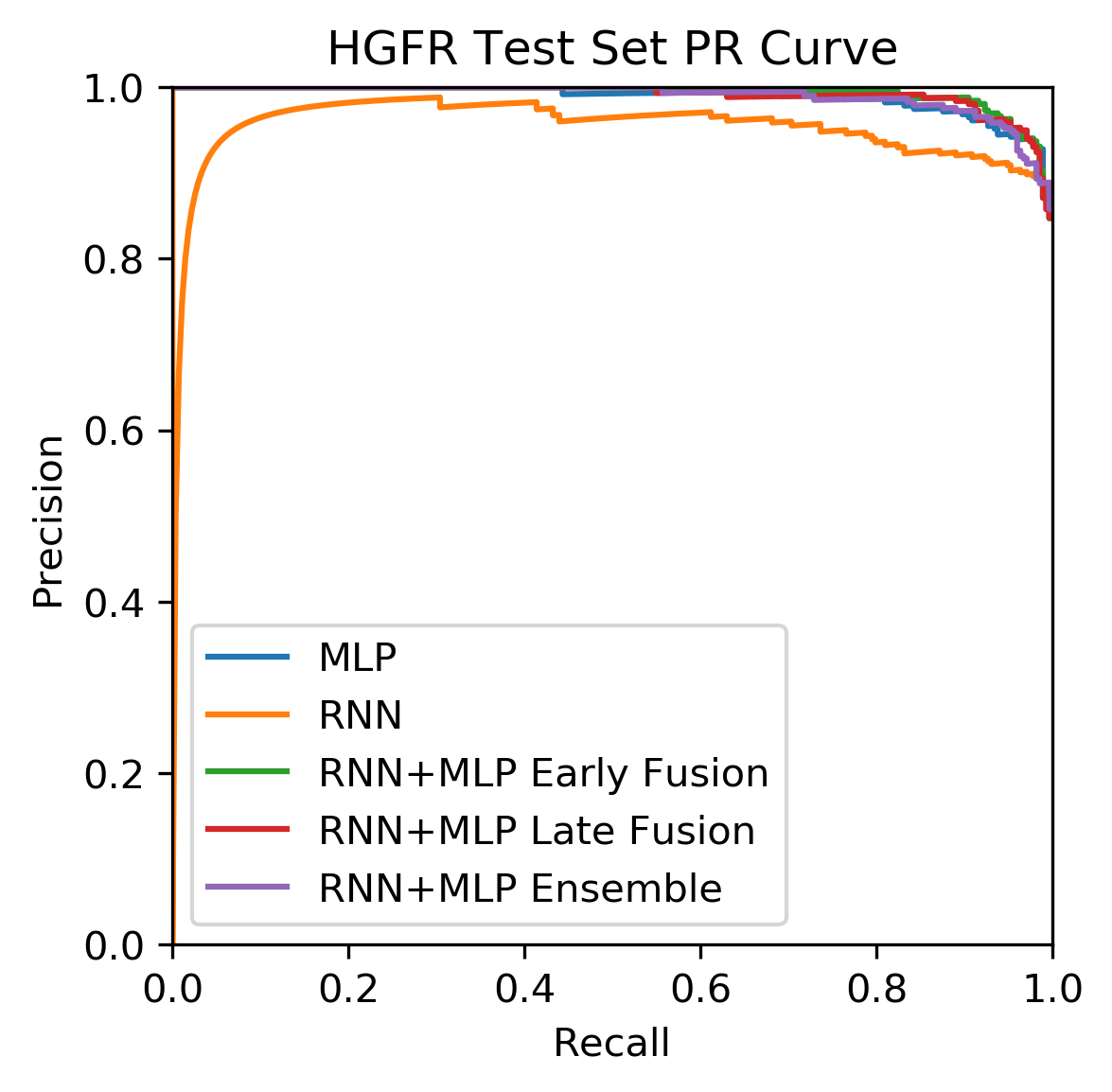}}
\end{minipage}\par\medskip
\begin{minipage}{.5\linewidth}
\centering
\subfloat[]{\label{fig:prc_e}\includegraphics[scale=.5]{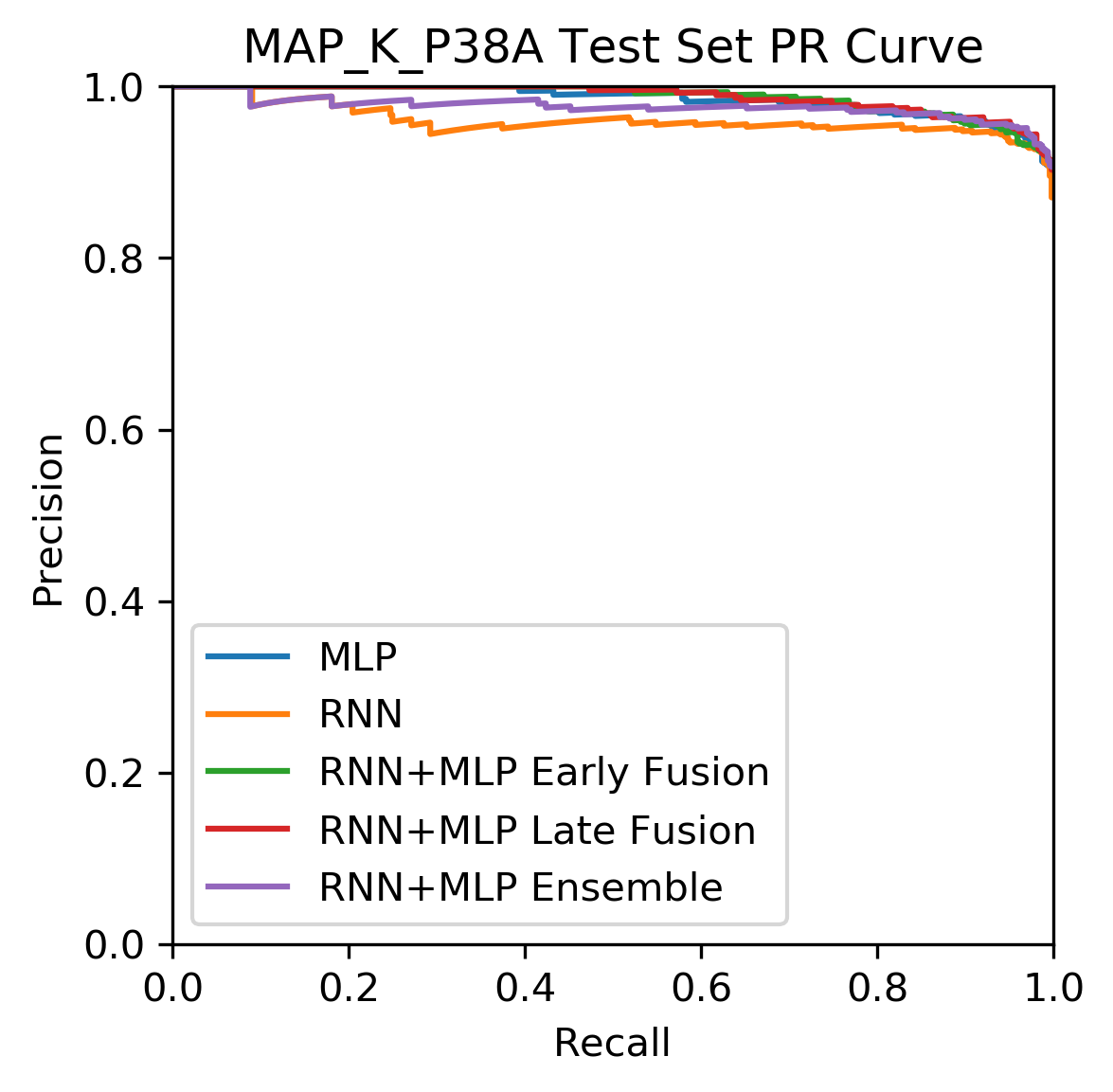}}
\end{minipage}%
\begin{minipage}{.5\linewidth}
\centering
\subfloat[]{\label{fig:prc_f}\includegraphics[scale=.5]{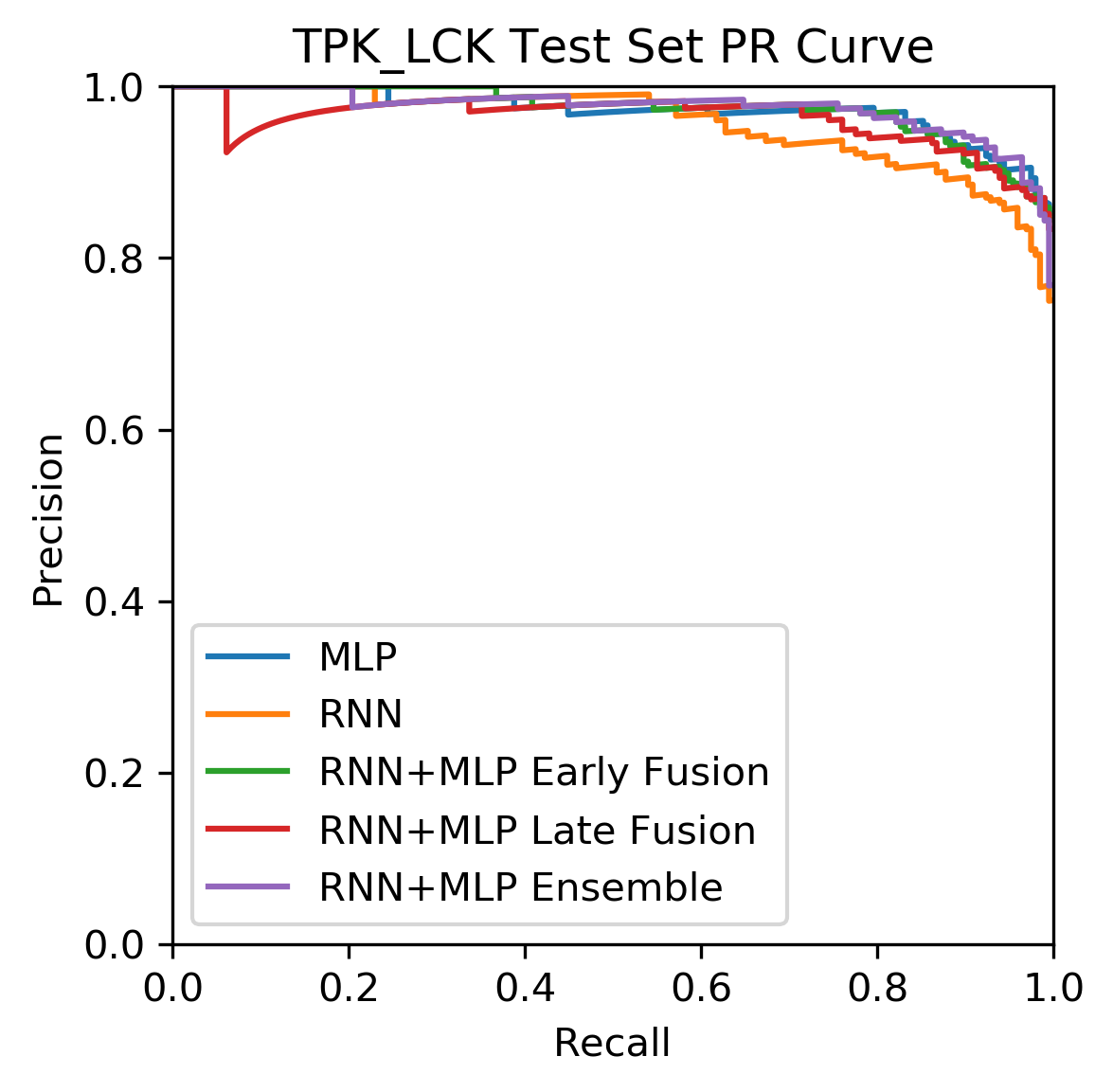}}
\end{minipage}\par\medskip
\begin{minipage}{.5\linewidth}
\centering
\subfloat[]{\label{fig:prc_g}\includegraphics[scale=.5]{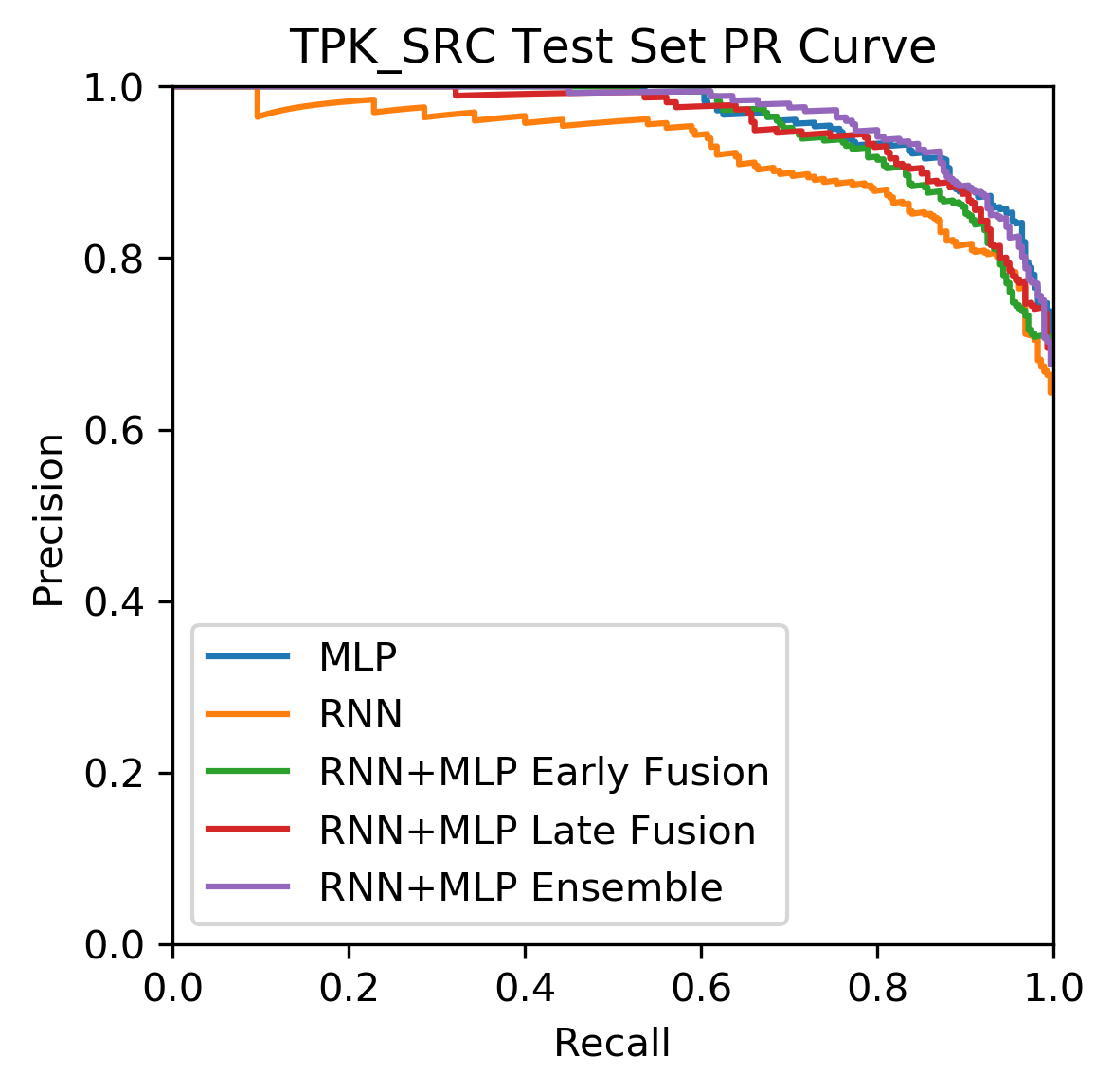}}
\end{minipage}%
\begin{minipage}{.5\linewidth}
\centering
\subfloat[]{\label{fig:prc_h}\includegraphics[scale=.5]{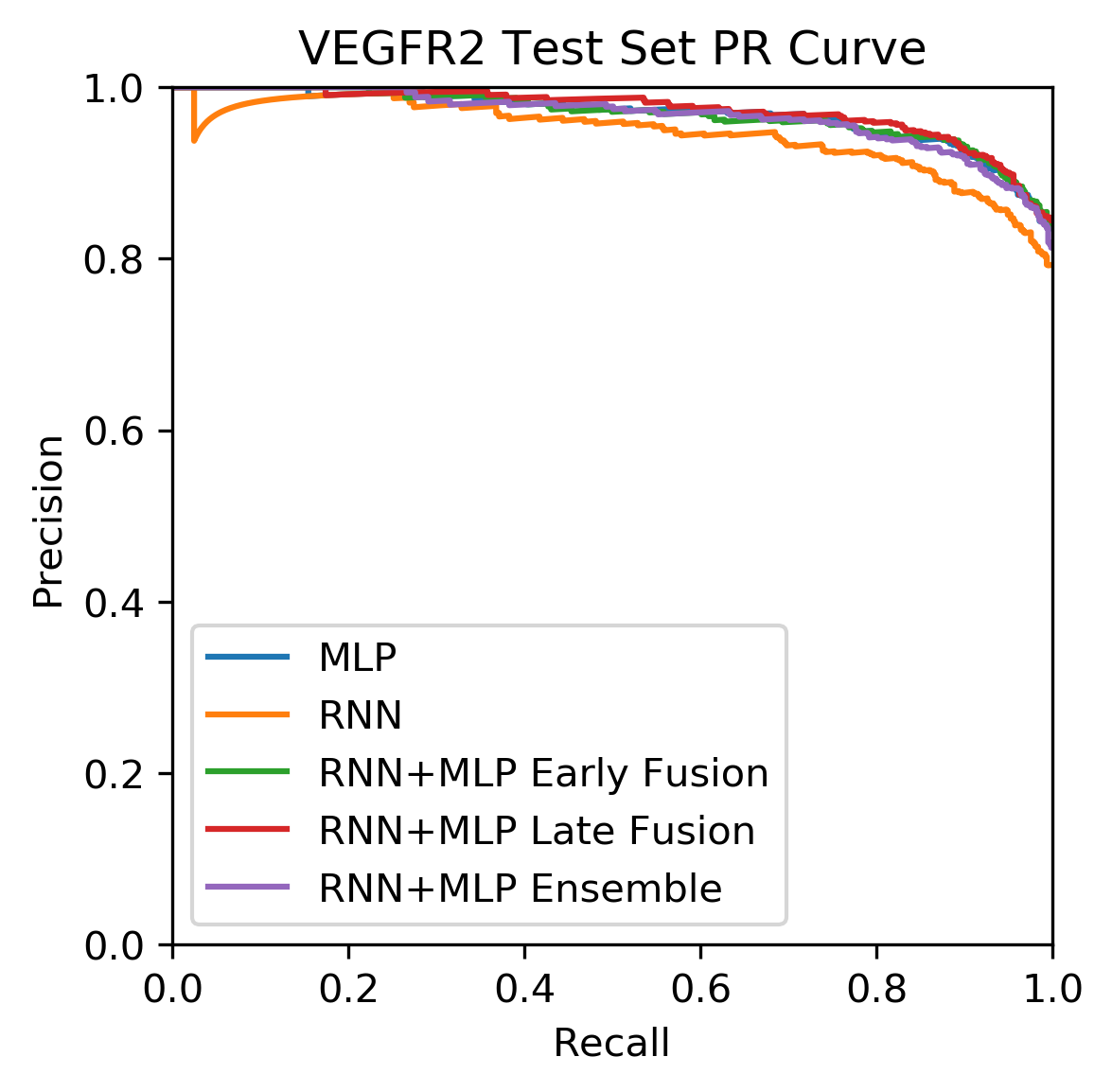}}
\end{minipage}
\caption{Testing set PRCs for each model for each of the 8 kinases using random splits.}
\label{fig:prc_random}
\end{figure}

\begin{figure}[htbp!]
\begin{minipage}{.5\linewidth}
\centering
\subfloat[]{\label{fig:croc_a}\includegraphics[scale=.5]{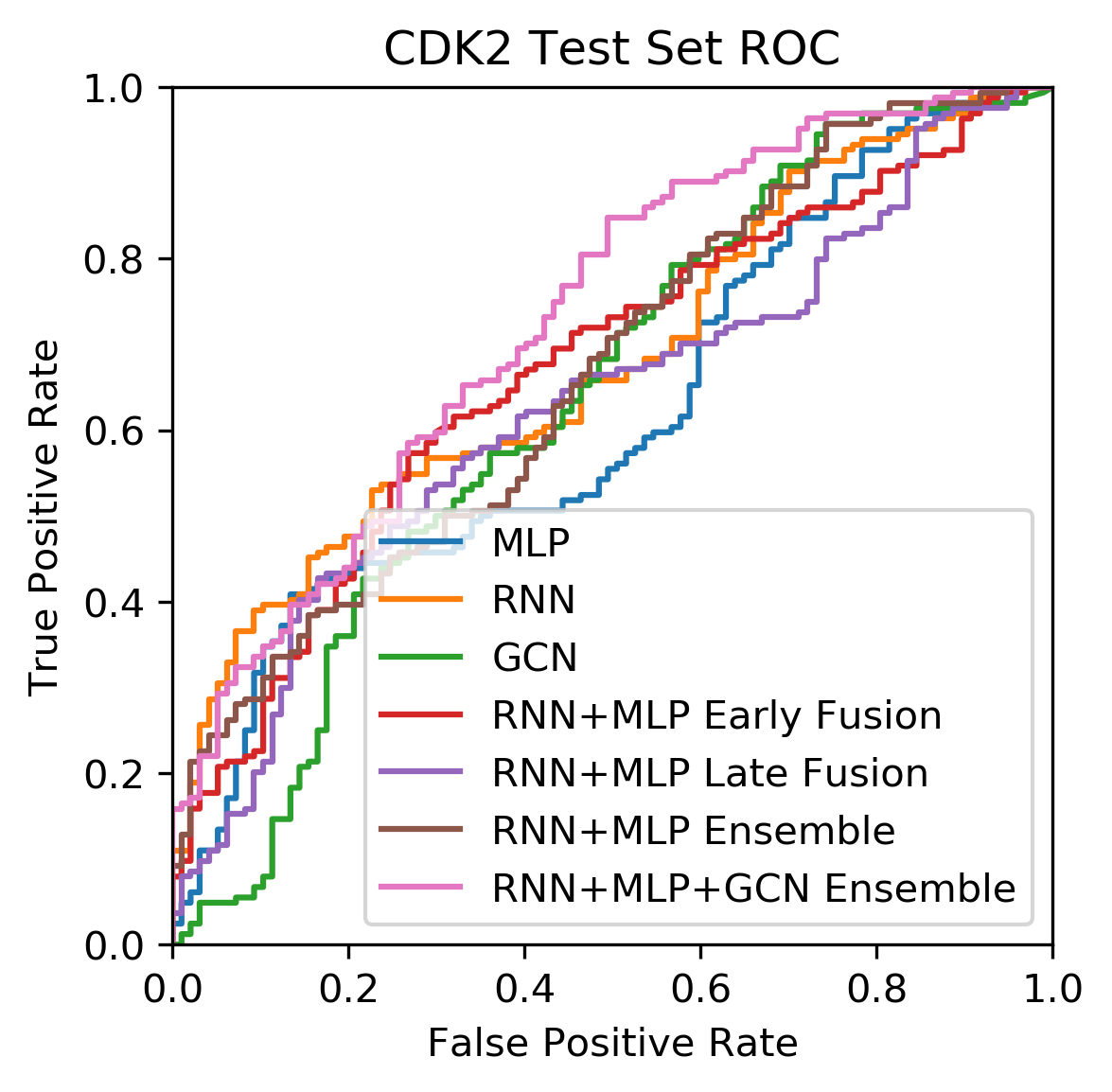}}
\end{minipage}%
\begin{minipage}{.5\linewidth}
\centering
\subfloat[]{\label{fig:croc_b}\includegraphics[scale=.5]{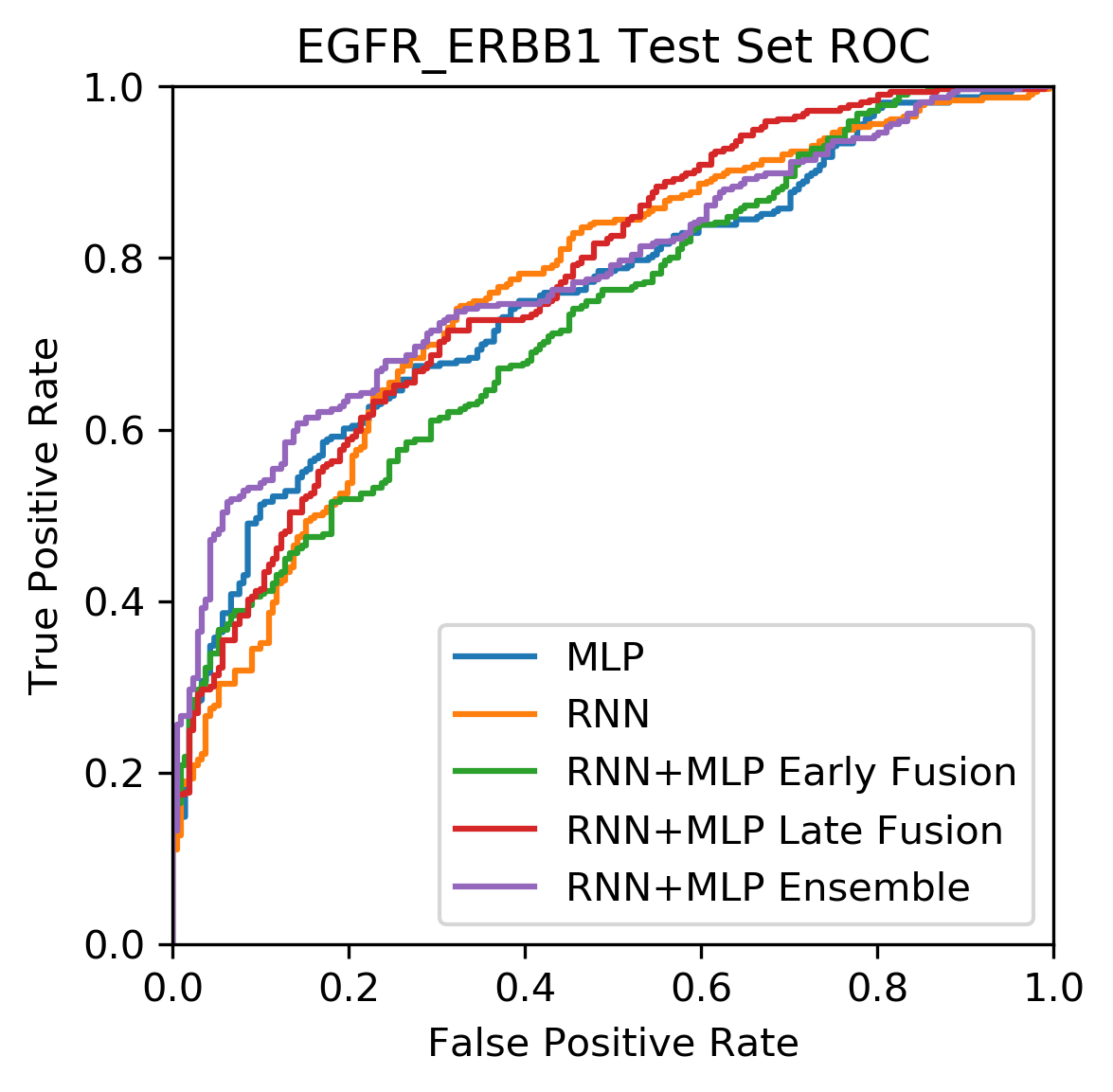}}
\end{minipage}\par\medskip
\centering
\begin{minipage}{.5\linewidth}
\centering
\subfloat[]{\label{fig:croc_c}\includegraphics[scale=.5]{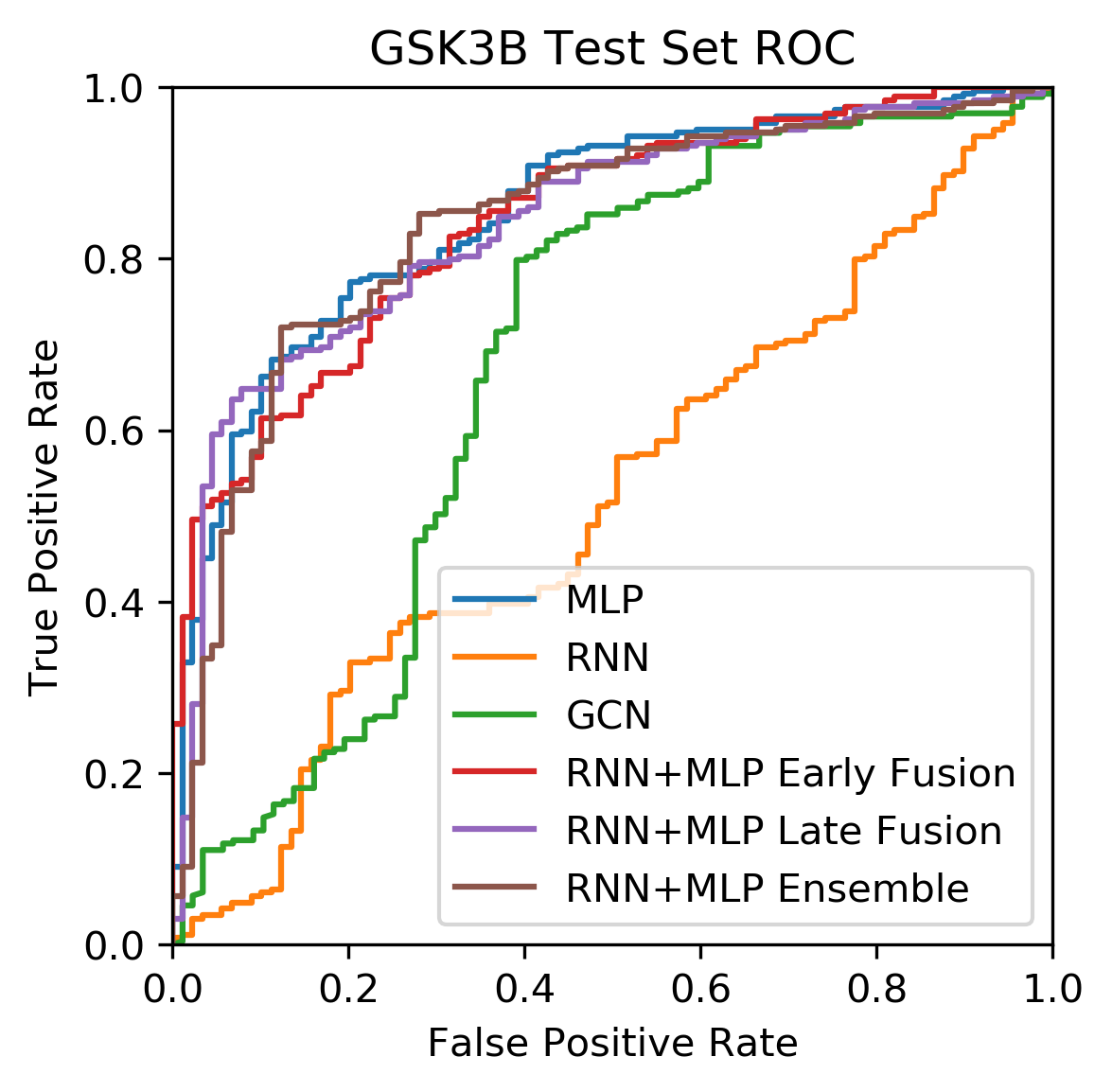}}
\end{minipage}%
\begin{minipage}{.5\linewidth}
\centering
\subfloat[]{\label{fig:croc_d}\includegraphics[scale=.5]{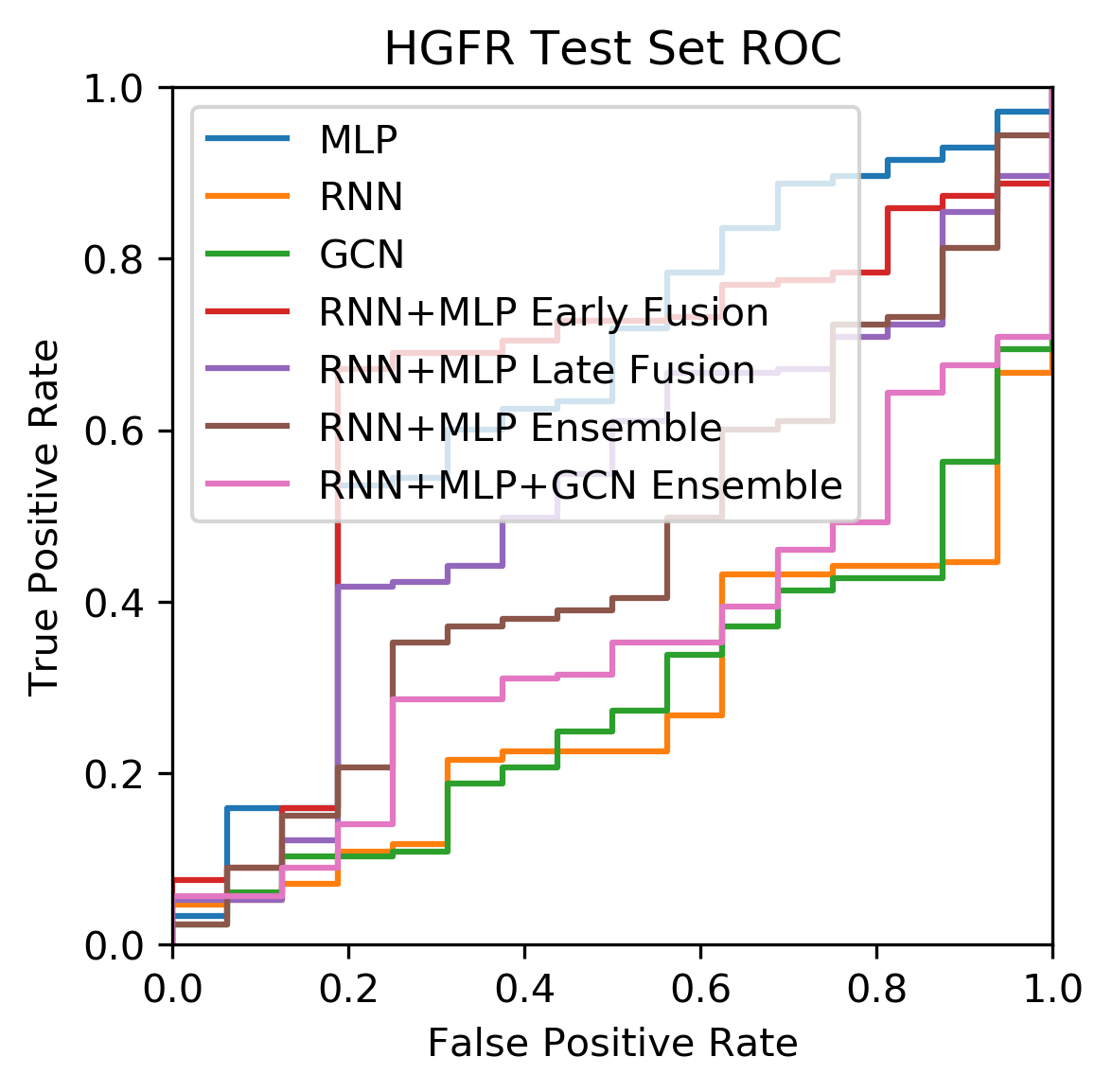}}
\end{minipage}\par\medskip
\begin{minipage}{.5\linewidth}
\centering
\subfloat[]{\label{fig:croc_e}\includegraphics[scale=.5]{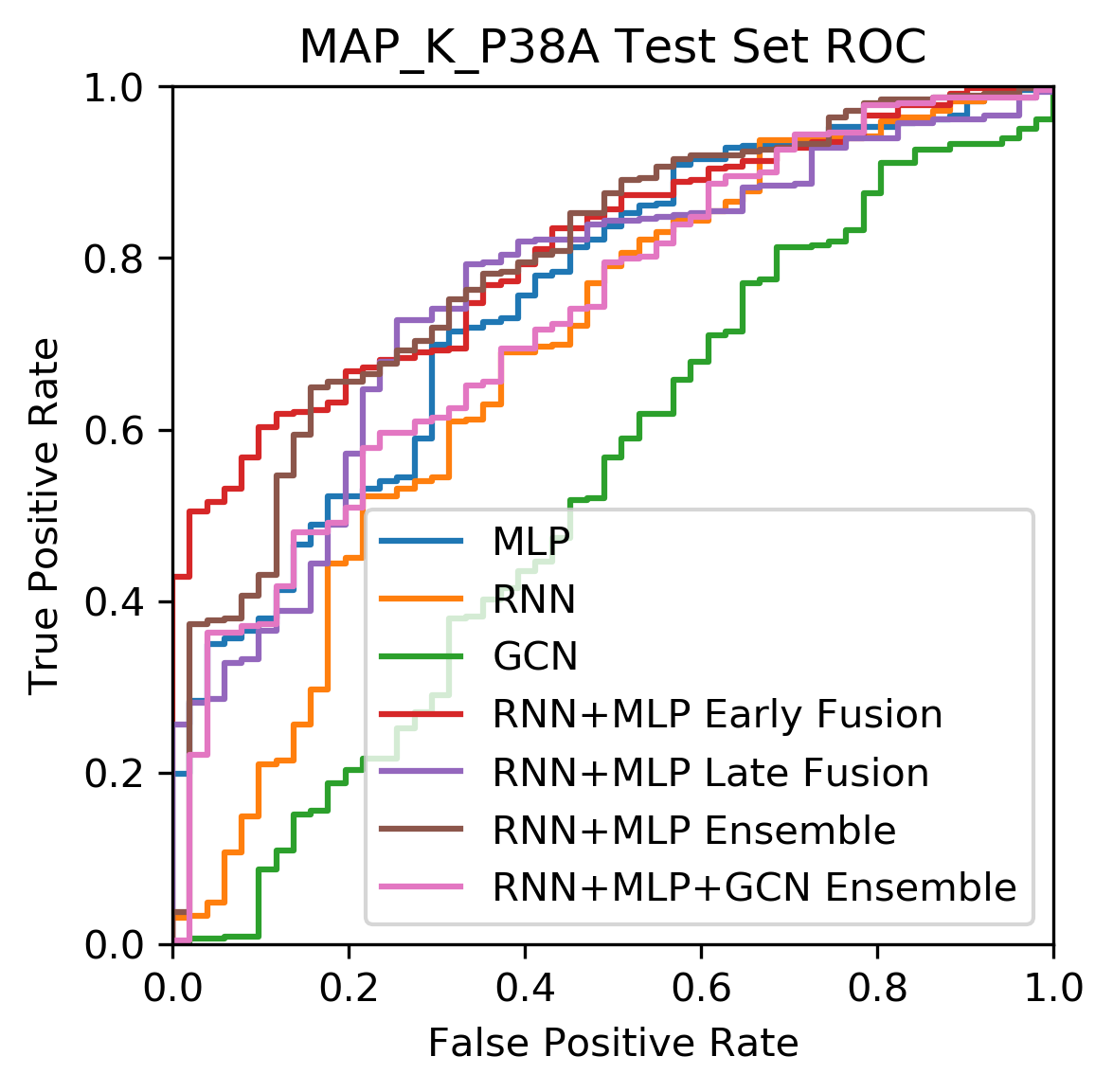}}
\end{minipage}%
\begin{minipage}{.5\linewidth}
\centering
\subfloat[]{\label{fig:croc_f}\includegraphics[scale=.5]{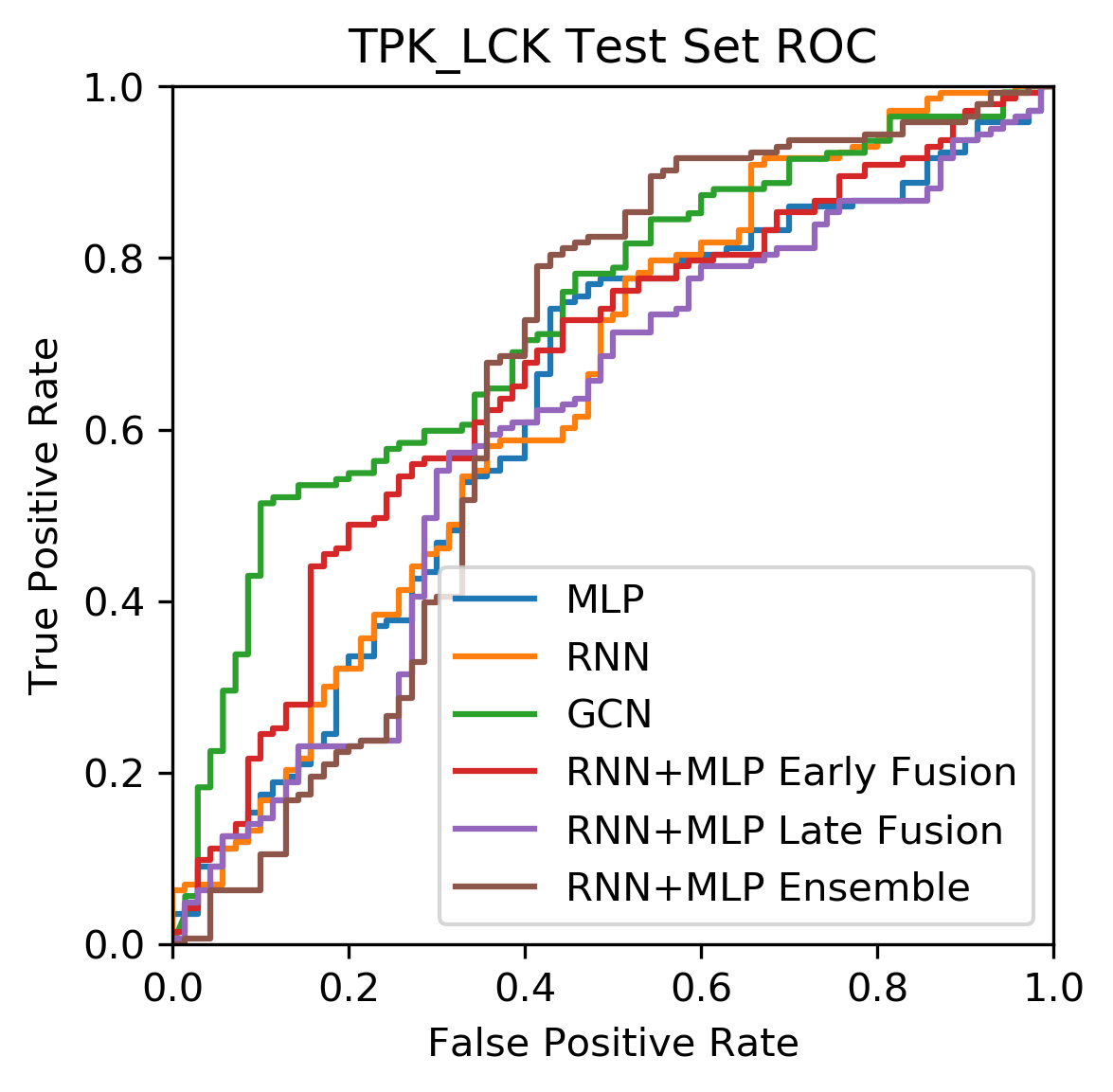}}
\end{minipage}\par\medskip
\begin{minipage}{.5\linewidth}
\centering
\subfloat[]{\label{fig:croc_g}\includegraphics[scale=.5]{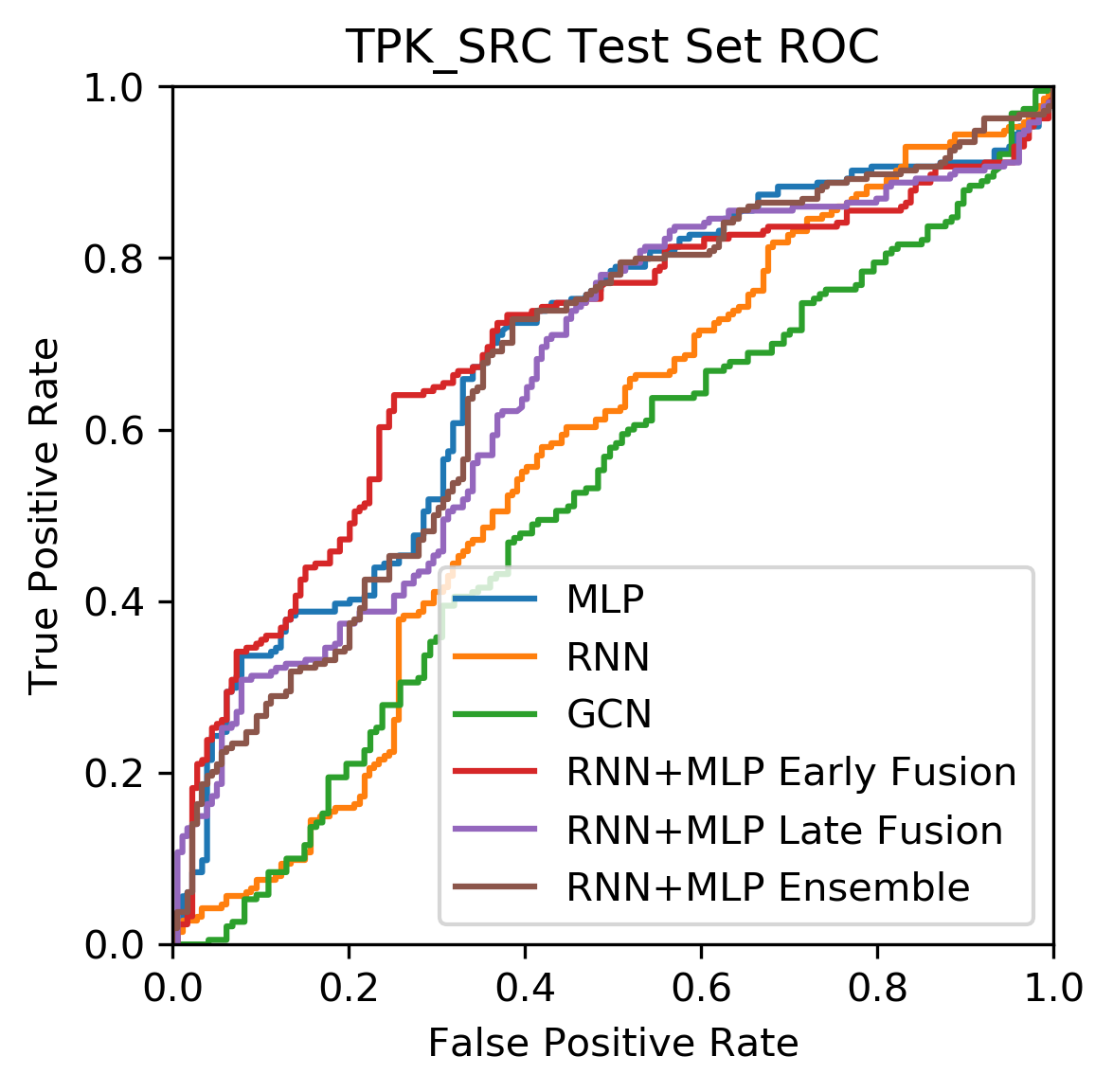}}
\end{minipage}%
\begin{minipage}{.5\linewidth}
\centering
\subfloat[]{\label{fig:croc_h}\includegraphics[scale=.5]{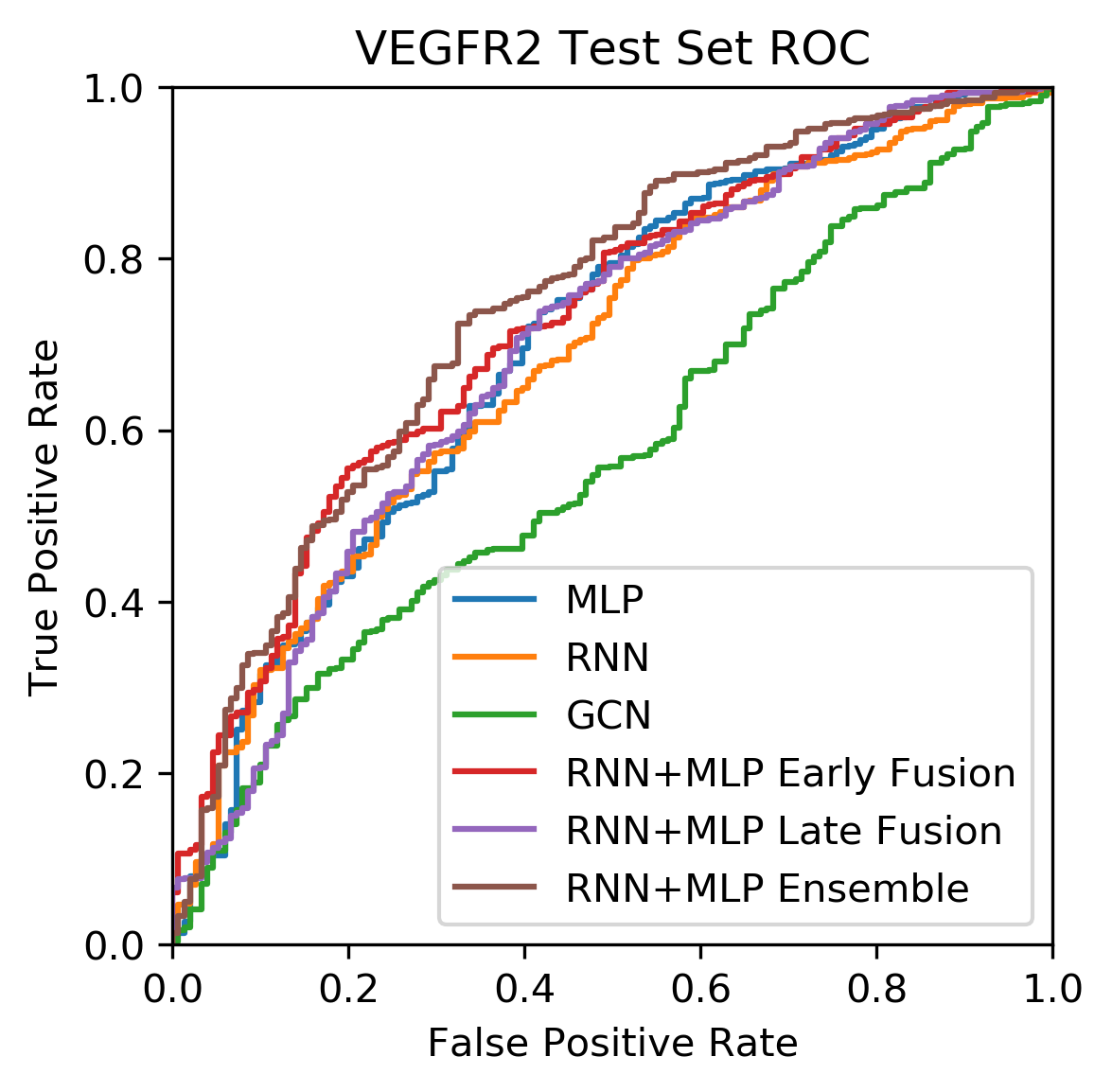}}
\end{minipage}
\caption{Testing set ROCs for each model for each of the 8 kinases using cluster-based splits.}
\label{fig:roc_cluster}
\end{figure}

\begin{figure}[htbp!]
\begin{minipage}{.5\linewidth}
\centering
\subfloat[]{\label{fig:cprc_a}\includegraphics[scale=.5]{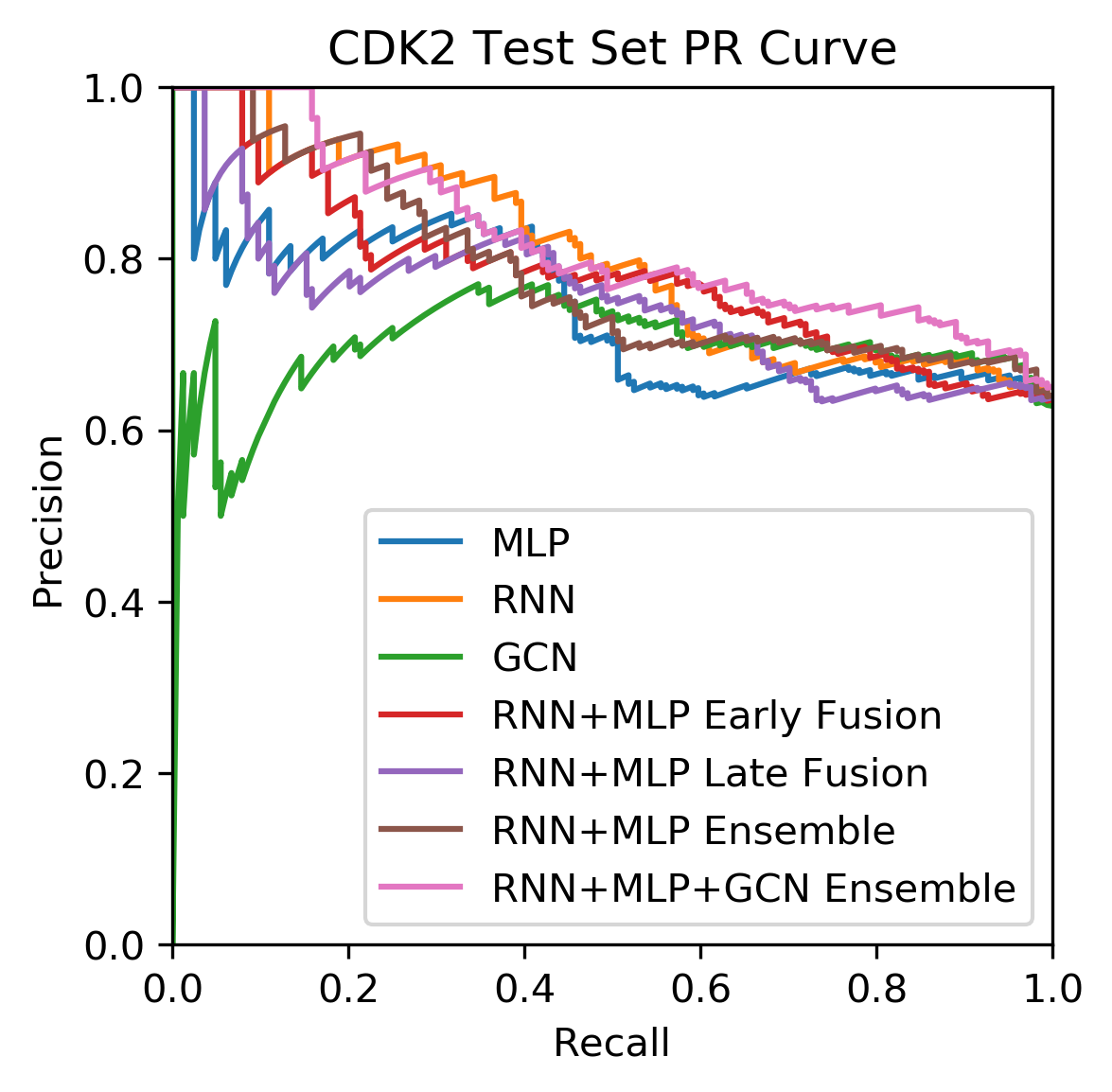}}
\end{minipage}%
\begin{minipage}{.5\linewidth}
\centering
\subfloat[]{\label{fig:cprc_b}\includegraphics[scale=.5]{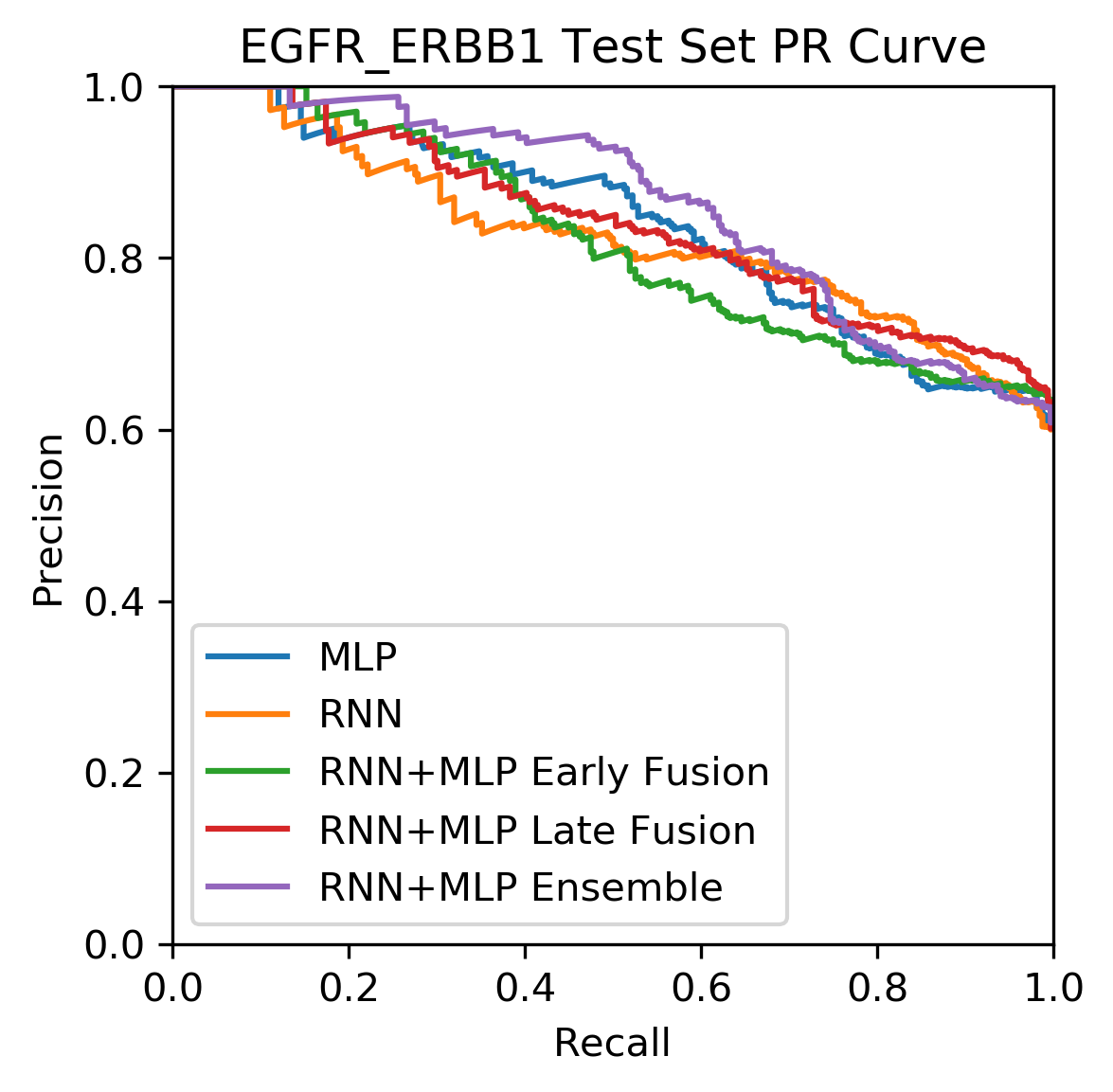}}
\end{minipage}\par\medskip
\centering
\begin{minipage}{.5\linewidth}
\centering
\subfloat[]{\label{fig:cprc_c}\includegraphics[scale=.5]{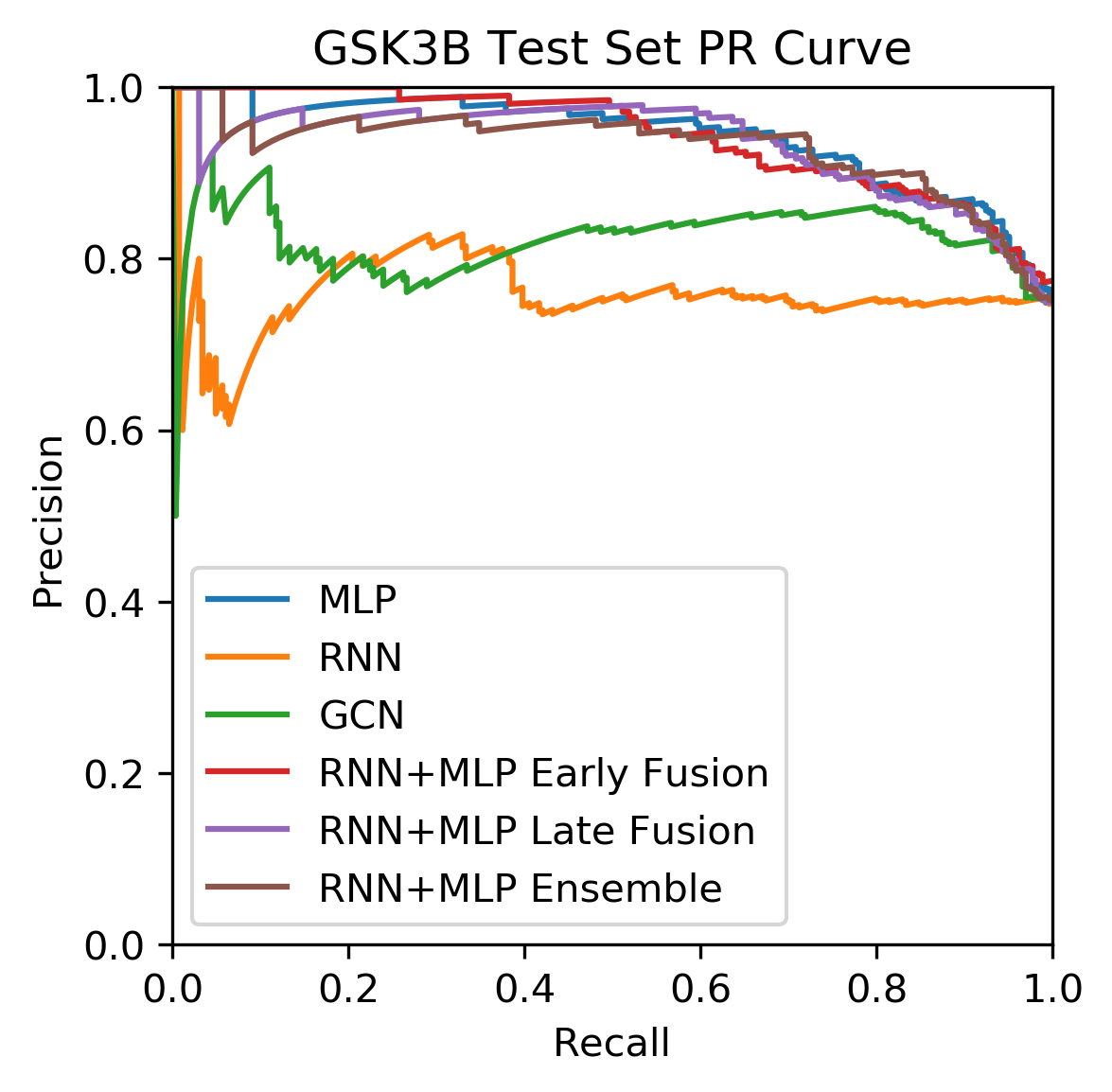}}
\end{minipage}%
\begin{minipage}{.5\linewidth}
\centering
\subfloat[]{\label{fig:cprc_d}\includegraphics[scale=.5]{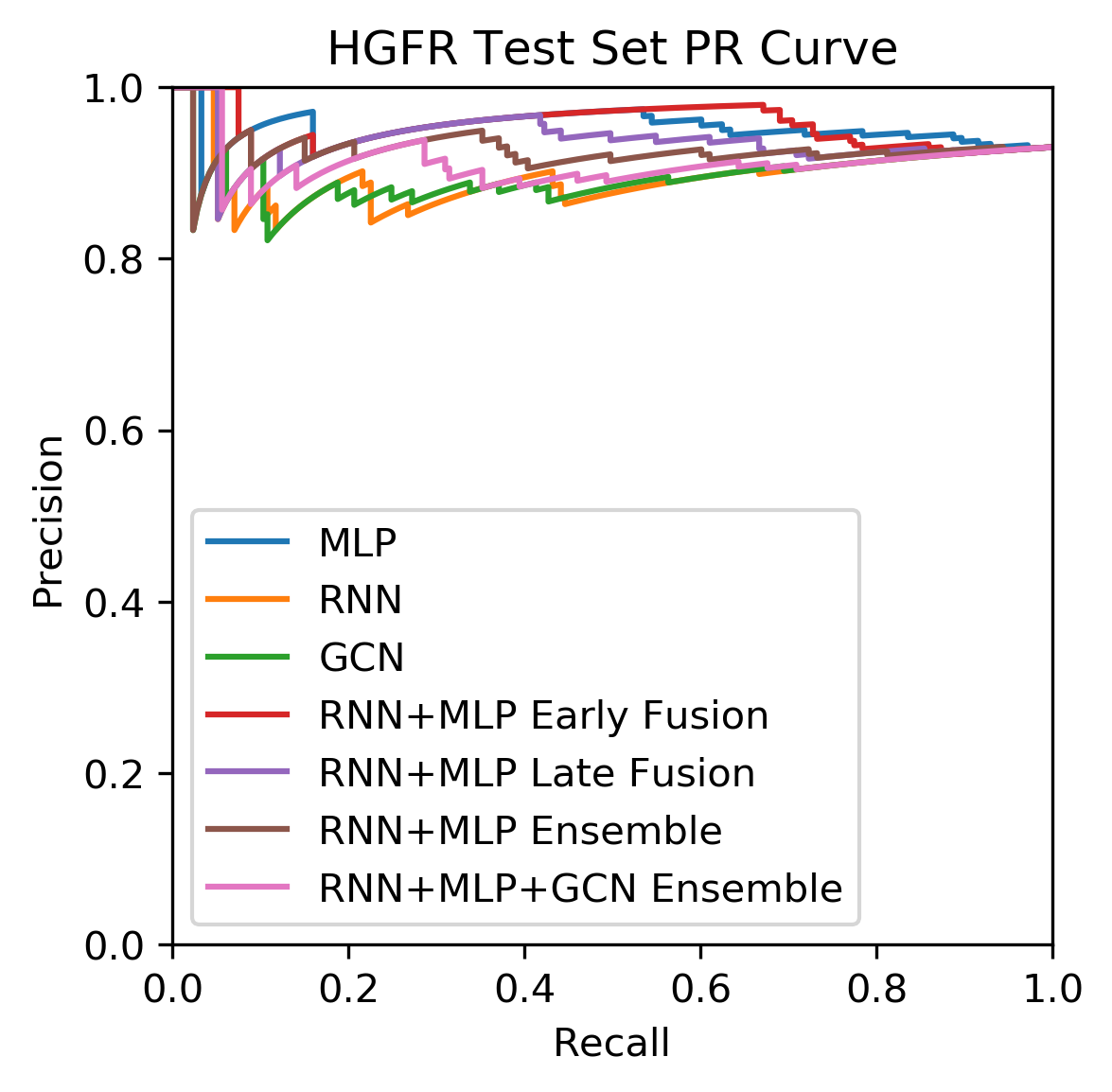}}
\end{minipage}\par\medskip
\begin{minipage}{.5\linewidth}
\centering
\subfloat[]{\label{fig:cprc_e}\includegraphics[scale=.5]{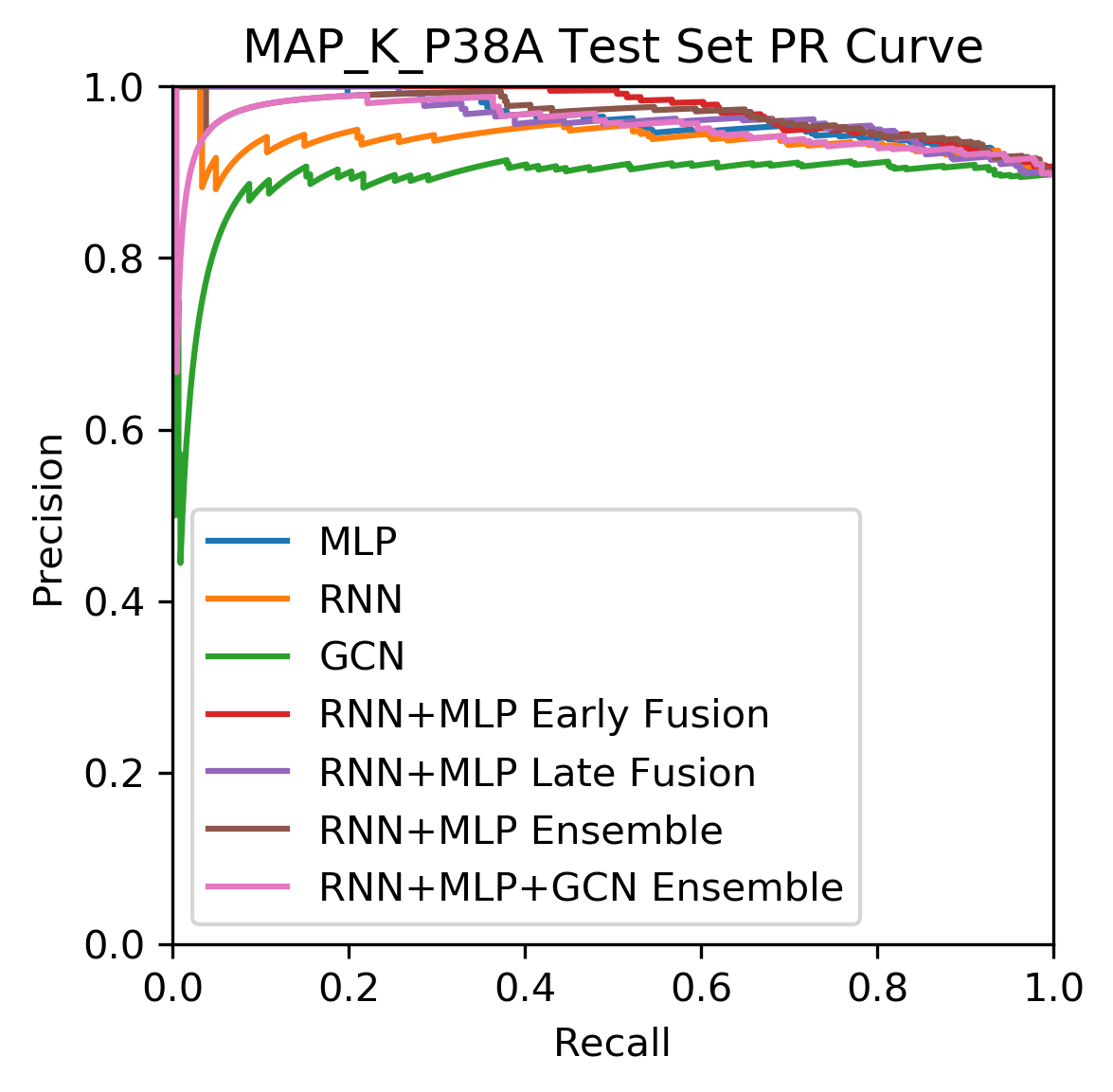}}
\end{minipage}%
\begin{minipage}{.5\linewidth}
\centering
\subfloat[]{\label{fig:cprc_f}\includegraphics[scale=.5]{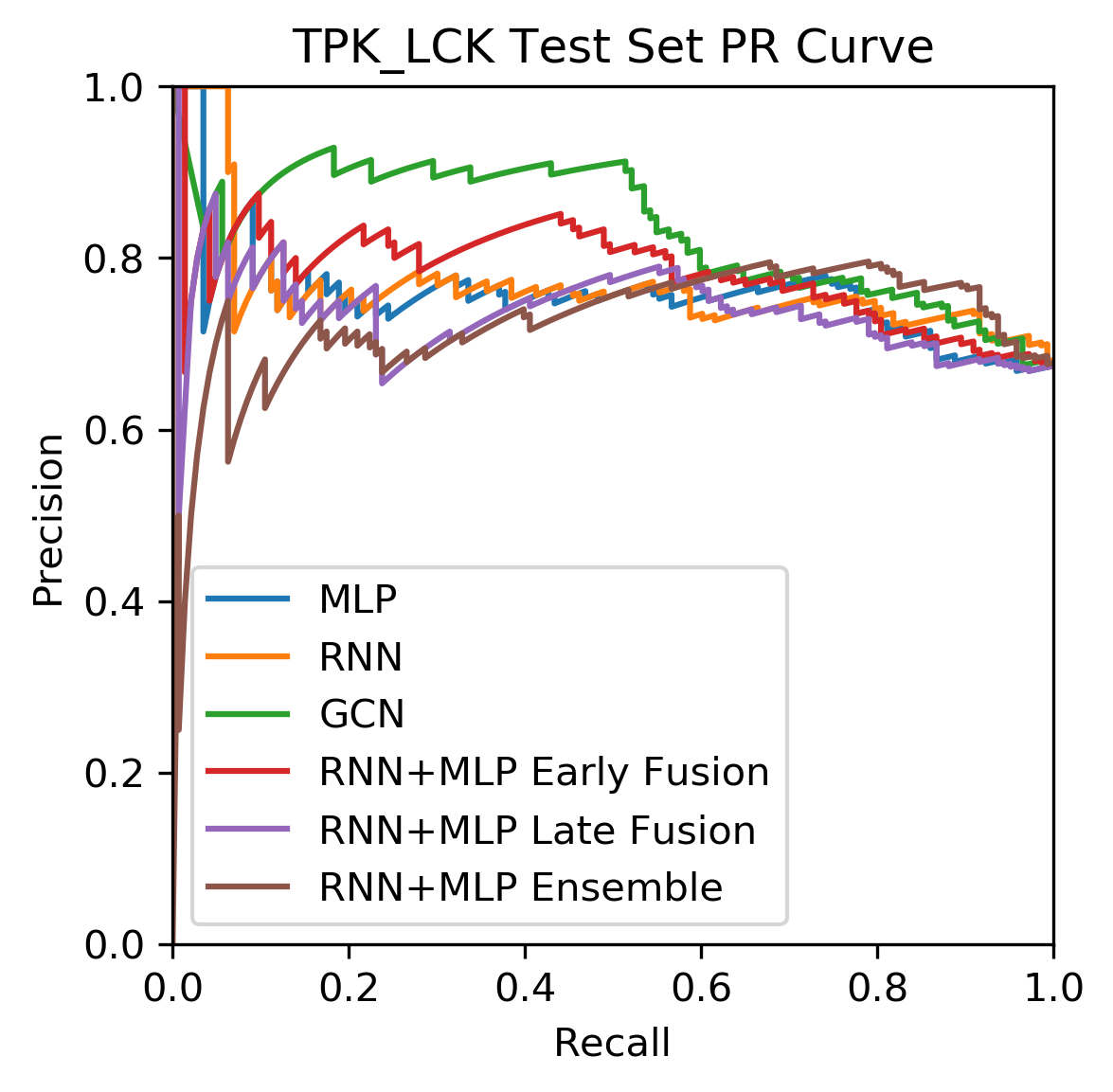}}
\end{minipage}\par\medskip
\begin{minipage}{.5\linewidth}
\centering
\subfloat[]{\label{fig:cprc_g}\includegraphics[scale=.5]{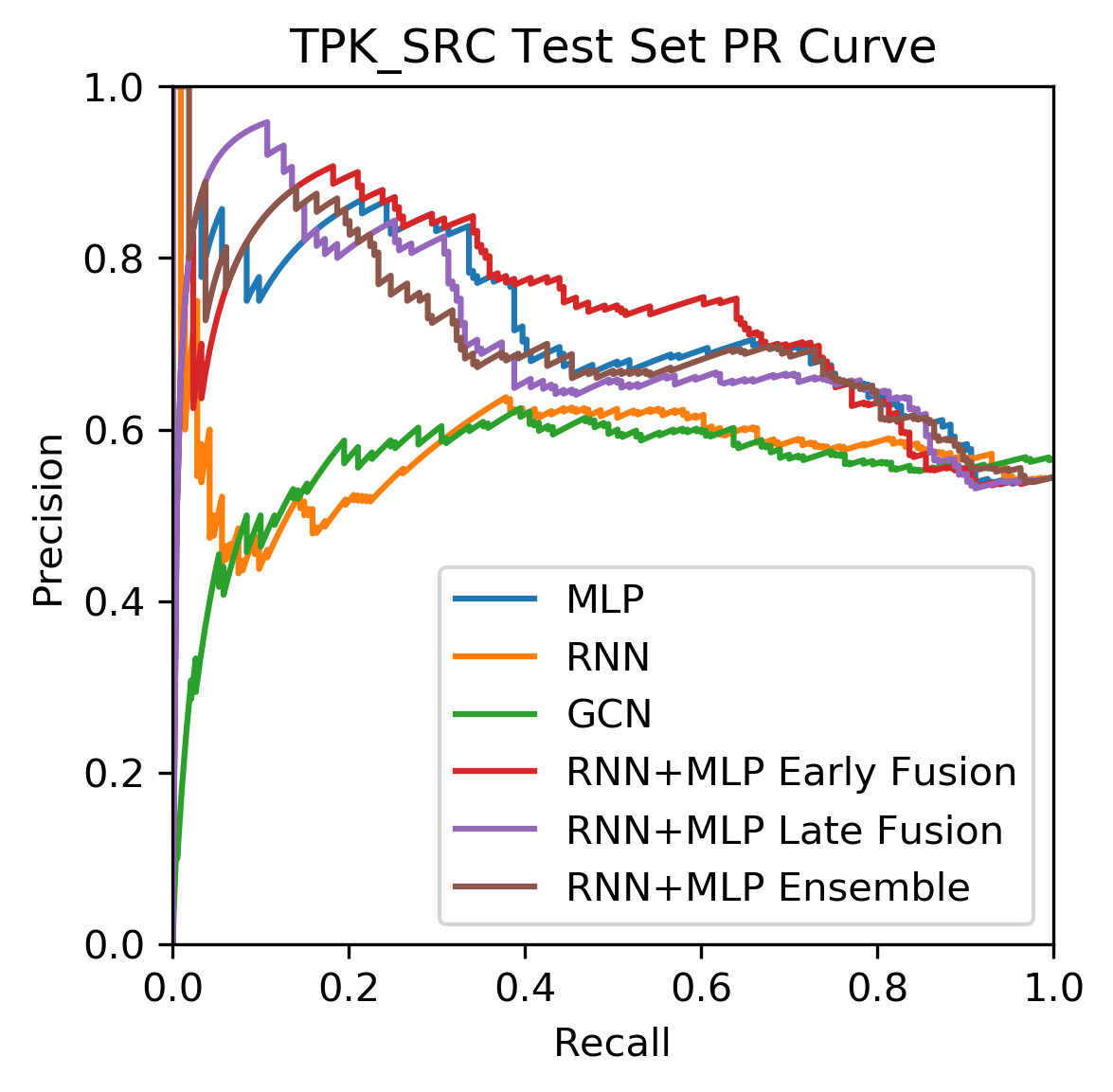}}
\end{minipage}%
\begin{minipage}{.5\linewidth}
\centering
\subfloat[]{\label{fig:cprc_h}\includegraphics[scale=.5]{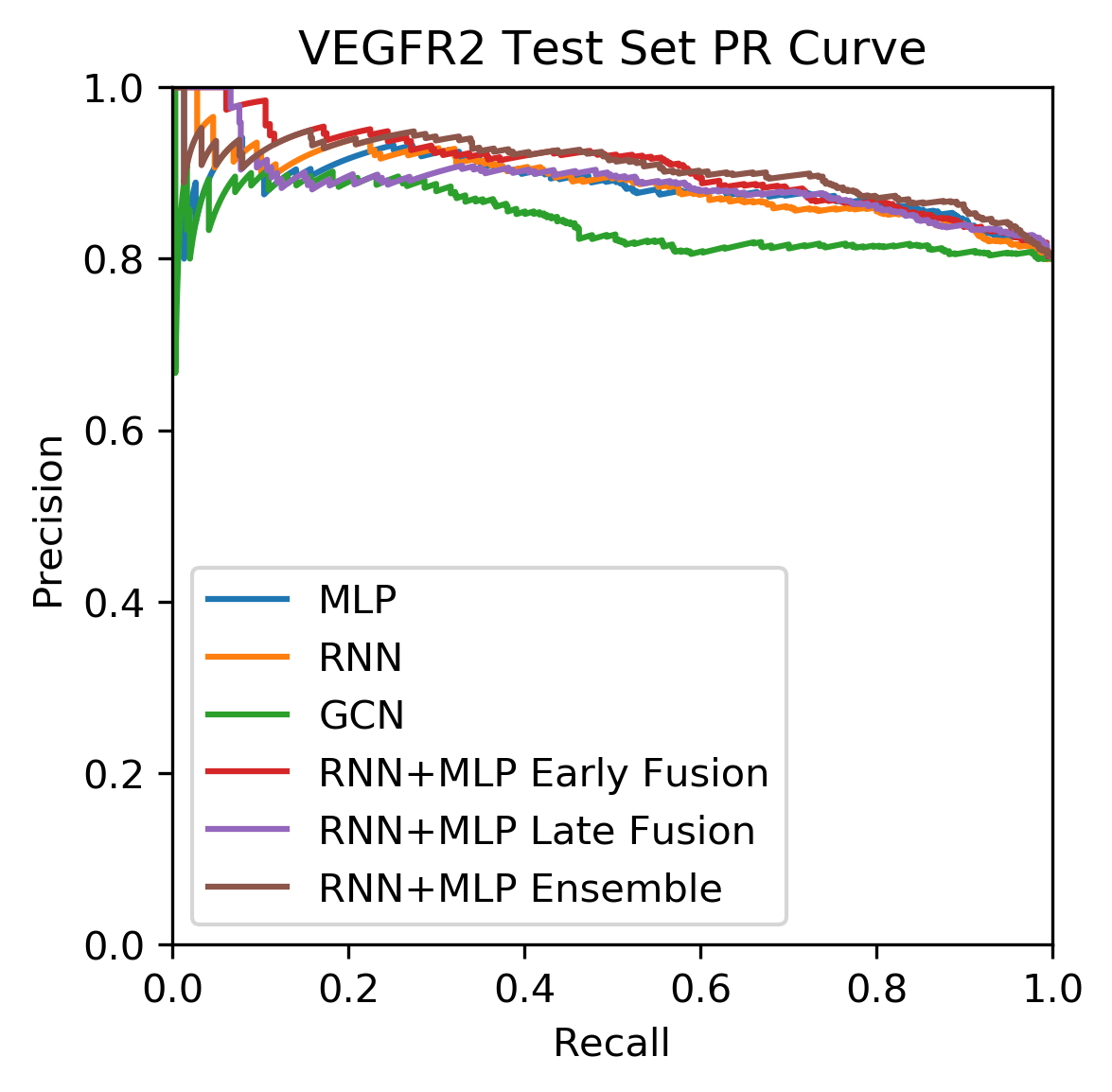}}
\end{minipage}
\caption{Testing set PRCs for each model for each of the 8 kinases using cluster-based splits.}
\label{fig:prc_cluster}
\end{figure}

We independently train each of the models for each of the kinases until the validation loss does not improve for 20 consecutive epochs. We then generate probability predictions from each of the trained models on the training, validation, and testing sets, and compute the AU-ROC and mean Average Precision (mAP). We use AU-ROC because this performance metric is robust to label imbalances, and we use mAP because in this context, we are particularly interested in the positive class (actual inhibitors that could serve as potential drugs). The GCN models were only trained and evaluated on the cluster-based splits. A GCN model was trained for each kinase except EGFR\textunderscore ERBB1 (insufficient training time). RNN+MLP+GCN ensemble predictions were computed and evaluated on the cluster-based test split for CDK2, HGFR, and MAP\textunderscore K\textunderscore P38A ( The performance of each model using random splits is included in Table \ref{table:results_random}, and the performance of each model using the cluster-based splits are included in Table \ref{table:results_cluster}.  Testing set ROCs and PRCs for each model using random splits are included in Figure \ref{fig:roc_random} and Figure \ref{fig:prc_random} respectively, and testing set ROCs and PRCs for each model using cluster-based splits are included in Figure \ref{fig:roc_cluster} and Figure \ref{fig:prc_cluster} respectively.
\begin{table}[htbp!]
\begin{center}
\begin{tabular}{|l|c|c|c|c|c|c|c|c|}
\hline
\textbf{Model} & \textbf{Train AUC} & \textbf{Train mAP} & \textbf{Val AUC} & \textbf{Val mAP} & \textbf{Test AUC} & \textbf{Test mAP} \\
\hline
\hline
\multicolumn{7}{|c|}{\textbf{CDK2}}\\
\hline
MLP & 0.976 & 0.985 & 0.931 & 0.951 & 0.931 & 0.961 \\
\hline
RNN & 0.946 & 0.967 & 0.869 & 0.899 & 0.896 & 0.932 \\
\hline
RNN+MLP Early Fusion & 0.923 & 0.951 & 0.915 & 0.936 & 0.903 & 0.942 \\
\hline
RNN+MLP Late Fusion & 0.977 & 0.987 & 0.919 & 0.941 & \textbf{0.937} & \textbf{0.965} \\
\hline
RNN+MLP Ensemble & 0.977 & 0.986 & 0.929 & 0.949 & 0.932 & 0.960 \\
\hline
\multicolumn{7}{|c|}{\textbf{EGFR\textunderscore ERBB1}}\\
\hline
MLP & 0.962 & 0.982 & 0.891 & 0.953 & 0.883 & 0.941 \\
\hline
RNN & 0.903 & 0.953 & 0.819 & 0.917 & 0.828 & 0.910 \\
\hline
RNN+MLP Early Fusion & 0.940 & 0.971 & 0.891 & 0.954 & 0.876 & 0.938 \\
\hline
RNN+MLP Late Fusion & 0.960 & 0.981 & 0.895 & 0.955 & \textbf{0.889} & \textbf{0.945} \\
\hline
RNN+MLP Ensemble & 0.956 & 0.979 & 0.887 & 0.950 & 0.883 & 0.939 \\
\hline
\multicolumn{7}{|c|}{\textbf{GSK3B}}\\
\hline
MLP & 0.983 & 0.993 & 0.922 & 0.967 & 0.909 & 0.968 \\
\hline
RNN & 0.920 & 0.965 & 0.833 & 0.902 & 0.794 & 0.900 \\
\hline
RNN+MLP Early Fusion & 0.976 & 0.990 & 0.908 & 0.965 & 0.894 & 0.962 \\
\hline
RNN+MLP Late Fusion & 0.990 & 0.996 & 0.934 & 0.975 & \textbf{0.917} & \textbf{0.971} \\
\hline
RNN+MLP Ensemble & 0.974 & 0.989 & 0.908 & 0.960 & 0.899 & 0.962 \\
\hline
\multicolumn{7}{|c|}{\textbf{HGFR}}\\
\hline
MLP & 0.978 & 0.996 & 0.955 & 0.990 & 0.943 & 0.989\\
\hline
RNN & 0.931 & 0.987 & 0.881 & 0.970 & 0.842 & 0.951 \\
\hline
RNN+MLP Early Fusion & 0.991 & 0.998 & 0.952 & 0.989 & \textbf{0.964} & \textbf{0.993}\\
\hline
RNN+MLP Late Fusion & 0.988 & 0.998 & 0.946 & 0.987 & 0.954 & 0.991 \\
\hline
RNN+MLP Ensemble & 0.981 & 0.997 & 0.953 & 0.990 & 0.946 & 0.990 \\
\hline
\multicolumn{7}{|c|}{\textbf{MAP\textunderscore K\textunderscore P38A}}\\
\hline
MLP & 0.985 & 0.997 & 0.913 & 0.981 & 0.923 & 0.985 \\
\hline
RNN & 0.959 & 0.992 & 0.820 & 0.963 & 0.844 & 0.961 \\
\hline
RNN+MLP Early Fusion & 0.974 & 0.995 & 0.915 & 0.982 & 0.931 & 0.988 \\
\hline
RNN+MLP Late Fusion & 0.972 & 0.995 & 0.917 & 0.984 & \textbf{0.934} & \textbf{0.988}\\
\hline
RNN+MLP Ensemble & 0.985 & 0.997 & 0.902 & 0.979 & 0.903 & 0.976 \\
\hline
\multicolumn{7}{|c|}{\textbf{TPK\textunderscore LCK}}\\
\hline
MLP & 0.974 & 0.987 & 0.913 & 0.964 & 0.942 & 0.972 \\
\hline
RNN & 0.916 & 0.962 & 0.860 & 0.935 & 0.898 & 0.957 \\
\hline
RNN+MLP Early Fusion & 0.976 & 0.990 & 0.912 & 0.961 & 0.941 & 0.974 \\
\hline
RNN+MLP Late Fusion & 0.965 & 0.985 & 0.914 & 0.965 & 0.929 & 0.961 \\
\hline
RNN+MLP Ensemble & 0.969 & 0.986 & 0.900 & 0.954 & \textbf{0.945} & \textbf{0.974} \\
\hline
\multicolumn{7}{|c|}{\textbf{TPK\textunderscore SRC}}\\
\hline
MLP & 0.987 & 0.992 & 0.945 & 0.970 & 0.941 & 0.966 \\
\hline
RNN & 0.910 & 0.944 & 0.839 & 0.910 & 0.885 & 0.924 \\
\hline
RNN+MLP Early Fusion & 0.973 & 0.985 & 0.943 & 0.969 & 0.920 & 0.956 \\
\hline
RNN+MLP Late Fusion & 0.963 & 0.977 & 0.938 & 0.966 & 0.926 & 0.957 \\
\hline
RNN+MLP Ensemble & 0.981 & 0.989 & 0.933 & 0.966 & \textbf{0.943} & \textbf{0.967}\\
\hline
\multicolumn{7}{|c|}{\textbf{VEGFR2}}\\
\hline
MLP & 0.969 & 0.992 & 0.893 & 0.969 & 0.900 & 0.968 \\
\hline
RNN & 0.943 & 0.985 & 0.839 & 0.943 & 0.836 & 0.946 \\
\hline
RNN+MLP Early Fusion & 0.984 & 0.996 & 0.900 & 0.971 & 0.899 & 0.968 \\
\hline
RNN+MLP Late Fusion & 0.971 & 0.992 & 0.900 & 0.971 & \textbf{0.911} & \textbf{0.973} \\
\hline
RNN+MLP Ensemble & 0.947 & 0.993 & 0.895 & 0.969 & 0.891 & 0.966 \\
\hline
\end{tabular}
\caption{Training, testing, and validation AU-ROC and mAP for each model for each of the 8 kinases using random splits. Best test set AUCs and mAPs bolded.}
\label{table:results_random}
\end{center}
\end{table}

\begin{table}[htbp!]
\small
\begin{center}
\begin{tabular}{|l|c|c|c|c|c|c|c|c|}
\hline
\textbf{Model} & \textbf{Train AUC} & \textbf{Train mAP} & \textbf{Val AUC} & \textbf{Val mAP} & \textbf{Test AUC} & \textbf{Test mAP} \\
\hline
\hline
\multicolumn{7}{|c|}{\textbf{CDK2}}\\
\hline
MLP & 0.937 & 0.961 & 0.586 & 0.716 & 0.619 & 0.742 \\
\hline
RNN & 0.860 & 0.914 & 0.694 & 0.846 & 0.679 & 0.803\\
\hline
GCN & 0.858 & 0.908 & 0.640 & 0.711 & 0.637 & 0.698 \\
\hline
RNN+MLP Early Fusion & 0.934 & 0.960 & 0.628 & 0.757 & 0.671 & 0.782 \\
\hline
RNN+MLP Late Fusion & 0.939 & 0.963 & 0.577 & 0.703 & 0.628 & 0.748 \\
\hline
RNN+MLP Ensemble & 0.934 & 0.958 & 0.681 & 0.760 & 0.666 & 0.784 \\
\hline
RNN+MLP+GCN Ensemble & 0.933 & 0.957 & 0.719 & 0.783 & \textbf{0.730} & \textbf{0.821} \\
\hline
\multicolumn{7}{|c|}{\textbf{EGFR\textunderscore ERBB1}}\\
\hline
MLP & 0.933 & 0.970 & 0.719 & 0.854 & 0.755 & 0.838 \\
\hline
RNN & 0.777 & 0.895 & 0.723 & 0.841 & 0.760 & 0.827 \\
\hline
RNN+MLP Early Fusion & 0.924 & 0.966 & 0.680 & 0.834 & 0.729 & 0.820 \\
\hline
RNN+MLP Late Fusion & 0.894 & 0.953 & 0.718 & 0.848 & 0.773 & 0.840 \\
\hline
RNN+MLP Ensemble & 0.920 & 0.962 & 0.737 & 0.864 & \textbf{0.779} & \textbf{0.862} \\
\hline
\multicolumn{7}{|c|}{\textbf{GSK3B}}\\
\hline
MLP & 0.964 & 0.987 & 0.815 & 0.916 & \textbf{0.853} & 0.943 \\
\hline
RNN & 0.719 & 0.877 & 0.573 & 0.766 & 0.523 & 0.759 \\
\hline
GCN & 0.717 & 0.864 & 0.571 & 0.709 & 0.677 & 0.824 \\
\hline 
RNN+MLP Early Fusion & 0.969 & 0.988 & 0.810 & 0.914 & 0.844 & \textbf{0.943} \\
\hline
RNN+MLP Late Fusion & 0.970 & 0.989 & 0.814 & 0.917 & 0.840 & 0.933 \\
\hline
RNN+MLP Ensemble & 0.959 & 0.984 & 0.810 & 0.898 & 0.839 & 0.930 \\
\hline
\multicolumn{7}{|c|}{\textbf{HGFR}}\\
\hline
MLP & 0.996 & 0.999 & 0.621 & 0.968 & \textbf{0.639} & 0.952 \\
\hline
RNN & 0.717 & 0.918 & 0.411 & 0.924 & 0.276 & 0.897 \\
\hline
GCN & 0.872 & 0.968 & 0.463 & 0.941 & 0.284 & 0.897 \\
\hline
RNN+MLP Early Fusion & 0.990 & 0.998 & 0.639 & 0.965 & 0.638 & \textbf{0.954} \\
\hline
RNN+MLP Late Fusion & 0.988 & 0.997 & 0.527 & 0.951 & 0.522 & 0.937 \\
\hline
RNN+MLP Ensemble & 0.992 & 0.998 & 0.483 & 0.938 & 0.455 & 0.927\\
\hline
RNN+MLP+GCN Ensemble & & & & & 0.351 & 0.913\\
\hline
\multicolumn{7}{|c|}{\textbf{MAP\textunderscore K\textunderscore P38A}}\\
\hline
MLP & 0.989 & 0.998 & 0.812 & 0.971 & 0.754 & 0.963 \\
\hline
RNN & 0.782 & 0.940 & 0.636 & 0.924 & 0.685 & 0.937 \\
\hline
GCN & 0.846 & 0.959 & 0.577 & 0.911 & 0.532 & 0.891 \\
\hline
RNN+MLP Early Fusion & 0.981 & 0.996 & 0.819 & 0.972 & \textbf{0.809} & \textbf{0.975} \\
\hline
RNN+MLP Late Fusion & 0.981 & 0.996 & 0.817 & 0.974 & 0.757 & 0.964 \\
\hline
RNN+MLP Ensemble & 0.982 & 0.996 & 0.821 & 0.966 & 0.793 & 0.967 \\
\hline
RNN+MLP+GCN Ensemble & & & & & 0.732 & 0.953\\
\hline
\multicolumn{7}{|c|}{\textbf{TPK\textunderscore LCK}}\\
\hline
MLP & 0.921 & 0.967 & 0.566 & 0.761 & 0.627 & 0.758 \\
\hline
RNN & 0.749 & 0.872 & 0.608 & 0.759 & 0.643 & 0.766 \\
\hline
GCN & 0.894 & 0.955 & 0.684 & 0.850 & \textbf{0.728} & \textbf{0.832} \\
\hline
RNN+MLP Early Fusion & 0.927 & 0.970 & 0.588 & 0.775 & 0.667 & 0.786 \\
\hline
RNN+MLP Late Fusion & 0.964 & 0.986 & 0.515 & 0.716 & 0.604 & 0.740 \\
\hline
RNN+MLP Ensemble & 0.915 & 0.964 & 0.602 & 0.727 & 0.654 & 0.729 \\
\hline
\multicolumn{7}{|c|}{\textbf{TPK\textunderscore SRC}}\\
\hline
MLP & 0.952 & 0.973 & 0.696 & 0.793 & 0.683 & 0.713\\
\hline
RNN & 0.765 & 0.861 & 0.573 & 0.671 & 0.567 & 0.580 \\
\hline
GCN & 0.763 & 0.841 & 0.518 & 0.636 & 0.518 & 0.559 \\
\hline
RNN+MLP Early Fusion & 0.921 & 0.954 & 0.677 & 0.791 & \textbf{0.694} & \textbf{0.732} \\
\hline
RNN+MLP Late Fusion & 0.922 & 0.953 & 0.646 & 0.763 & 0.657 & 0.704 \\
\hline
RNN+MLP Ensemble & 0.928 & 0.956 & 0.668 & 0.778 & 0.670 & 0.705\\
\hline
\multicolumn{7}{|c|}{\textbf{VEGFR2}}\\
\hline
MLP & 0.963 & 0.990 & 0.711 & 0.908 & 0.700 & 0.886 \\
\hline
RNN & 0.839 & 0.951 & 0.698 & 0.893 & 0.687 & 0.888 \\
\hline
GCN & 0.640 & 0.863 & 0.514 & 0.821 & 0.574 & 0.843 \\
\hline
RNN+MLP Early Fusion & 0.962 & 0.990 & 0.723 & 0.918 & 0.724 & \textbf{0.907} \\
\hline
RNN+MLP Late Fusion & 0.960 & 0.989 & 0.717 & 0.909 & 0.696 & 0.888 \\
\hline
RNN+MLP Ensemble & 0.953 & 0.987 & 0.754 & 0.923 & \textbf{0.744} & 0.906 \\
\hline
\end{tabular}
\caption{Training, testing, and validation AU-ROC and mAP for each model for each of the 8 kinases using cluster-based splits. Best test set AUCs and mAPs bolded.}
\label{table:results_cluster}
\end{center}
\end{table}

\section{Conclusion}
In conclusion, we were able to utilize various neural networks to classify small molecules as inhibiting/non-inhibiting with the random split. For the initial random training-validation-testing split, the RNN+MLP late fusion model performed the best out of all the algorithms that we tried, with a test AUROC of $>0.9$. As expected, the performance decreased with the harsher cluster-based splits. Surprisingly, on the  cluster-based splits, early fusion and RNN+MLP ensembles, rather than RNN+MLP late fusion, resulted in the best performance. Even with the harsher cluster-based splits, the best models yielded AUROCs of $>0.7$ and mAPs of $>0.8$, indicating the models still learn chemically relevant information to accurately make predictions rather than just learn to classify based on similar structures from the training set. Moreover, the graph convolution network underperformed compared to the MLP on the fingerprints, which suggests that the GCN requires further hyperparameter tuning to match or exceed state of the art methods. However, for TPK\textunderscore LCK the GCN was the best model, and for CDK2, the RNN+MLP+GCN model was the best model, so this shows promise that with further optimization, GCNs can be powerful. Fused models performs the best, so it is a promising approach to leverage multiple feature representations to optimize classification performance.

\subsection{Future Work}
First, we can examine the specific chemical features learned by the model by taking model parameter gradients over features to identify salient features, as well as the biological features of the kinases that may account for the difference in performance across the kinases. Second, we could also explore different fingerprints and feature vectors for the MLP and RNN implementation, since different fingerprints can often have varying amounts of success in different situations. Next, we could start including the vast amount of unlabeled data from both Kaggle and the CHemBL dataset to begin semi-supervised learning. Because of the sheer volume of potential small molecules, and the relative rarity of labeled inhibitors, semi-supervised learning could have immense benefits.  In addition, the CHemBL database has IC50 value information, which means that rather than performing binary classification, we could explore a regression task, trying to predict IC50 values.  Finally, we could examine the structures of well-performing molecules to determine any commonalities, which could lead to insight into potential inhibitor classes.

\subsection{Incorporating Feedback}
We clarified our specific usage of Kaggle dataset, as requested by the peer reviews. Specifically, we mention we only use the ChemBL IDs and labels from the Kaggle dataset, and extract all other information from other sources, namely SMILES strings from the ChemBL database and fingerprints using the Python RDKit package. We also clarified our procedures for generating train/validation/test splits, and included a more detailed description of SMILES codes as well as how the 1-hot vector embeddings for the RNN were generated. Moreover, in response to concerns that because we randomly split the data, various molecular substructures/structural similarities may be shared between our train, validation, and test sets, thus artificially inflating our validation and test accuracy, we implemented an additional cluster-based approach to split the data. The fusion models involve pretraining and augmentation, as suggested by one of the reviews. Moreover, we highlight and discuss trends with regards to which models perform the best across all the kinases in the conclusion, and leave to future work some of the suggestions to analyze specific biological features of the kinases or chemical features of the molecules that may contribute to performance differences. In addition, we elaborated on previous work done in the realm of inhibitor prediction, and clarified the Briem study. We have also given statistics for numbers of positively and negatively labeled molecules for each of the kinases in the kaggle dataset as suggested by reviewers.  Finally, we now include both ROCs (and AUROC) and Precision-Recall curves (and mAP) as visualizations and metrics as suggested by reviewers.

\subsection{Acknowledgments}
We would like to thank Dr. James Zou for his mentorship throughout this project. We would also like to thank Dr. Anshul Kundaje, Avanti Shrikumar, and Michael Wainberg for their teaching and feedback.

\bibliographystyle{ieeetr}
\bibliography{bibliography.bib}

\begin{thebibliography}{10}

\bibitem{siegal2018statistics}
R.~L. Siegal, K.~D. Miller, and A.~Jemal, ``Cancer statistics, 2018,'' in {\em
  CA: A Cancer Journal for Clinicians}, pp.~68:7--30, 2018.

\bibitem{huang2017effects}
C.-Y. Huang, D.-T. Ju, C.-F. Chang, P.~M. Reddy, and B.~K. Velmurugan, ``A
  review on the effects of current chemotherapy drugs and natural agents in
  treating non–small cell lung cancer,'' in {\em BioMedicine}, pp.~12--23,
  2017.

\bibitem{kannaiyan}
R.~Kannaiyan and D.~Mahadevan, ``A comprehensive review of protein kinase
  inhibitors for cancer therapy,'' in {\em Expert Review of Anticancer
  Therapy}, pp.~1249--1270, 2018.

\bibitem{Mishra}
R.~Mishra, ``Glycogen synthase kinase 3 beta: can it be a target for oral
  cancer,'' in {\em Molecular Cancer}, 2010.

\bibitem{cecchi}
F.~Cecchi, D.~C. Rabe, and D.~P. Bottaro, ``The hepatocyte growth factor
  receptor: Structure, function and pharmacological targeting in cancer,'' in
  {\em Molecular Cancer}, p.~146–151, 2010.

\bibitem{olson2004}
J.~M. Olson and A.~R. Hallahan, ``p38 map kinase: a convergence point in cancer
  therapy,'' in {\em TRENDS in Molecular Medicine}, pp.~125--129, 2010.

\bibitem{smith2010}
N.~R. Smith, D.~Baker, N.~H. James, K.~Ratcliffe, M.~Jenkins, S.~E. Ashton,
  G.~Sproat, R.~Swann, N.~Gray, A.~Ryan, J.~M. Jurgensmeier, and C.~Womack,
  ``Vascular endothelial growth factor receptors vegfr-2 and vegfr-3 are
  localized primarily to the vasculature in human primary solid cancers,'' in
  {\em Human Cancer Biology}, pp.~3548--61, 2010.

\bibitem{zhang2009inhibitors}
J.~Zhang, P.~L. Yang, and N.~S. Gray, ``Targeting cancer with small molecule
  kinase inhibitors,'' in {\em Nature Reviews}, pp.~28--39, 2009.

\bibitem{briem}
H.~Briem and J.~Gunther, ``Classifying “kinase inhibitor‐likeness” by
  using machine‐learning methods,'' in {\em Nature Reviews}, pp.~558--566,
  2005.

\bibitem{tsar}
``Tsar datasheet version 3.3 accelrys inc,'' in {\em
  http://www.3dsbiovia.com/products/datasheets/tsar.pdf}.

\bibitem{graph_conv_intro}
S.~Kearnes, K.~McCloskey, M.~Berndl, V.~Pande, and P.~Riley, ``Molecular graph
  convolutions: Moving beyond fingerprints,'' {\em arXiv preprint
  arXiv:1603.00856}, 2016.

\bibitem{kaggle}
K.~Xiao, ``Cancer inhibitors dataset.''
  \url{https://www.kaggle.com/xiaotawkaggle/inhibitors}.

\bibitem{smiles}
D.~Weininger, A.~Weininger, and J.~L. Weining, ``Smiles. 2. algorithm for
  generation of unique smiles notation,'' {\em Journal of Chemical Information
  and Modeling}, 1989.

\bibitem{atom_pair}
R.~E. Carhart, D.~H. Smith, and R.~Venkataraghavan, ``Atom pairs as molecular
  features in structure-activity studies: definition and applications,'' {\em
  Journal of Chemical Information and Computer Sciences}, 1985.

\bibitem{ecfp}
D.~Rogers and M.~Hahn, ``Extended-connectivity fingerprints,'' {\em Journal of
  Chemical Information and Modeling}, 2010.

\bibitem{topological_torsion}
R.~Nilakantan, N.~Bauman, J.~S. Dixon, and R.~Venkataraghavan, ``Topological
  torsion: a new molecular descriptor for sar applications. comparison with
  other descriptors,'' {\em Journal of Chemical Information and Computer
  Sciences}, 1987.

\bibitem{kingma2014adam}
D.~P. Kingma and J.~Ba, ``Adam: A method for stochastic optimization,'' {\em
  arXiv preprint arXiv:1412.6980}, 2014.

\bibitem{hochreiter1997long}
S.~Hochreiter and J.~Schmidhuber, ``Long short-term memory,'' {\em Neural
  computation}, vol.~9, no.~8, pp.~1735--1780, 1997.

\bibitem{chemi-net}
K.~Liu, X.~Sun, L.~Jia, J.~Ma, H.~Xing, J.~Wu, H.~Gao, Y.~Sun, F.~Boulnois, and
  J.~Fan, ``Chemi-net: A molecular graph convolutional network for accurate
  drug property prediction,'' {\em arXiv preprint arXiv:1803.06236}, 2017.

\end{thebibliography}

\end{document}